\documentclass{aa}

\def\h1{$h^{-1}$\SP}

\def\void#1{{}}

\def\j{$J$}
\def\h{$H$}
\def\k{$K_s$}

\usepackage{graphicx}
\usepackage{times}
\usepackage{amsmath}
\usepackage{amssymb}
\usepackage{mathptm}
\usepackage{graphicx}
\usepackage{psfig}
\usepackage{natbib}
\begin{document}


\title{ESO Imaging Survey: Infrared Deep Public Survey \thanks{Based
on observations collected at the European Southern Observatory, La
Silla, Chile within ESO programs 164.O-O561 and
169.A-725.}}

\author{ L.~F. Olsen\inst{1,2,3} \and J.-M. Miralles \inst{3,4} \and
  L. da Costa\inst{3,5}\and R. Madejsky\inst{6} \and
  H. E. J{\o}rgensen\inst{1} \and A. Mignano\inst{3,7,8} \and
  S. Arnouts\inst{9} \and C. Benoist\inst{2,3} \and
  J.~P. Dietrich\inst{10} \and R. Slijkhuis\inst{3} \and
  S. Zaggia\inst{11,12} }
\institute{
Dark Cosmology Centre, Niels Bohr Institute, University of Copenhagen, Juliane Maries Vej 30, 2100 Copenhagen,
Denmark
\and Observatoire de la C\^{o}te d'Azur, BP 4229, 06304 NICE
cedex 4, France
\and
European Southern Observatory, Karl-Schwarzschild-Str. 2, 85748
Garching b. M\"unchen, Germany
\and T\`ecniques d'Avantguarda, Avda. Carlemany 75, AD-700 Les Escaldes, Andorra\and
Observat\'orio Nacional, Rua Gen. Jos\'e Cristino 77, Rio de
Janeiro, R.J., Brasil
\and
Universidade Estadual de Feira de Santana, Campus Universit\'{a}rio, Feira de Santana, BA, Brazil
\and
Dipartimento di Astronomia, Universita' di Bologna, via Ranzani~1, 40126
Bologna, Italy
\and INAF - Istituto di Radioastronomia, Via Gobetti~101, 40129 Bologna,
Italy
\and Laboratoire d'astrophysique de Marseille, Traverse du Siphon, BP 8, 13376 Marseille, Cedex 12, France
\and Argelander-Institut f\"ur Astronomie\thanks{Founded by
    merging of the Institut f\"ur Astrophysik und Extraterrestrische
    Forschung, the Sternwarte, and the Radioastronomisches Institut
    der Universit\"at Bonn.}, University of Bonn, Auf dem H\"ugel 71,
    53121 Bonn, Germany
\and INAF - Osservatorio Astronomico di Trieste, via G.B. Tiepolo, 11, 34131, Trieste, Italy
\and INAF - Astronomical Observatory of Padova, Vicolo dell'Osservatorio 5, 35122 Padova, Italy
}

\offprints{Lisbeth Fogh Olsen, lisbeth@astro.ku.dk}

\date{Received ; accepted}

\abstract{This paper is part of the series presenting the final
results obtained by the ESO Imaging Survey (EIS) project. It presents
new \j\ and \k\ data obtained from observations conducted at the ESO
3.5m New Technology Telescope (NTT) using the SOFI camera.  These data
were taken as part of the Deep Public Survey (DPS) carried out by the
ESO Imaging Survey program, significantly extending the earlier
optical/infrared EIS-DEEP survey presented in a previous paper of this
series. The DPS-IR survey comprises two observing strategies: shallow
\k\ observations providing nearly full coverage of pointings with
complementary multi-band (in general $UBVRI$) optical data obtained
using ESO's wide-field imager (WFI) and deeper \j\ and \k\
observations of the central parts of these fields.  Currently, the
DPS-IR survey provides a coverage of roughly 2.1~square~degrees
($\sim300$~SOFI pointings) in \k\ with 0.63~square~degrees to fainter
magnitudes and also covered in \j, over three independent regions of
the sky. The goal of the present paper is to briefly describe the
observations, the data reduction procedures, and to present the final
survey products which include fully calibrated pixel-maps and catalogs
extracted from them. The astrometric solution with an estimated
accuracy of $\lesssim0\farcs15$ is based on the USNO catalog and
limited only by the accuracy of the reference catalog. The final
stacked images presented here number 89 and 272, in \j\ and \k,
respectively, the latter reflecting the larger surveyed area. The \j\
and \k\ images were taken with a median seeing of $0\farcs77$ and
$0\farcs8$. The images reach a median $5\sigma$ limiting magnitude of
$J_{AB}\sim23.06$ as measured within an aperture of 2\arcsec, while
the corresponding limiting magnitude in \k$_{AB}$ is $\sim21.41$ and
$\sim$22.16~mag for the shallow and deep strategies.  Although some
spatial variation due to varying observing conditions is observed,
overall the observed limiting magnitudes are consistent with those
originally proposed. The quality of the data has been assessed by
comparing the measured magnitude of sources at the bright end directly
with those reported by the 2MASS survey and at the faint end by
comparing the counts of galaxies and stars with those of other surveys
to comparable depth and to model predictions. The final science-grade
catalogs together with the astrometrically and photometrically
calibrated co-added images are available at CDS\thanks{Available at
CDS via anonymous ftp to cdsarc.u-strasbg.fr (130.79.128.5) or via
http://cdsweb.u-strasbg.fr/cgi-bin/qcat?J/A+A/}.

\keywords{catalogs -- surveys-- stars: general - galaxies: general} }

\maketitle

\section{Introduction}
\label{sec:intro}

Deep multi-wavelength observations of selected regions of the sky from
space and ground-based facilities, combined with spectroscopic
observations from large-aperture telescopes, offer the most promising
means to probe the distant universe and study in a comprehensive way
the evolution of galaxies and large-scale structures over a broad
interval of look-back time.  Foreseeing the need to complement
observations being carried out with the HST, Chandra, XMM, Spitzer and
GALEX telescopes with ground-based multi-passband data in optical and
infrared, the Working Group for public surveys (SWG) at ESO
recommended the ESO Imaging Survey (EIS, Renzini \& da Costa, 1997)
project to undertake deep, optical/infrared observations. 

This paper is part of a series describing recent data releases
from the ESO Imaging Survey using the newly implemented EIS Data
Reduction System \citep{dacosta04}. The reduction of optical data and
details of the released products are described in \citet[ hereafter
Paper~I]{dietrich05} in the context of the optical follow-up of the
XMM-Newton Serendipitous Sky Survey. In \citet[ hereafter
Paper~II]{olsen06a} the reduction of infrared data is described based
on the infrared part of the EIS-DEEP survey. The last survey discussed
in the present series is the EIS Deep Public Survey which combines
optical \citep[][ hereafter Paper~III]{mignano06} and infrared
observations, with the latter being the topic of the present paper.

The original infrared observations, conducted in 1998, targeted the
HDF-S and CDF-S regions and fully calibrated images, source catalogs
and high-redshift galaxy candidates were immediately made public prior
to the Science Verification of the first unit of the VLT \citep[see
preprints by ][]{dacosta98, rengelink98}. In spite of some
shortcomings in the original reduction of the infrared images (see
Paper~II for a more detailed discussion) these were used in
combination with the optical data, taken with the SUSI2 detector, to
produce color catalogs and to color-select high-redshift galaxy
candidates. These candidates were later used in spectroscopic
observations, conducted during commissioning of the VLT and the FORS
spectrograph \citep[e.g., ][]{Cristiani00}, which, in general,
confirmed their estimated photometric redshifts. However, the total
area of the survey was small and, in particular, only a small fraction
of the field of view of Chandra was covered by the available data.

In order to expand the original infrared coverage of EIS,
complementing other ongoing optical surveys, and provide full coverage
of the CDF-S $\lesssim350$~square~arcmin field and of the flanking
regions, the SWG designed the following strategy for the Deep Public
Survey (DPS). The EIS DPS survey consists of two parts: first, a deep,
optical multi-passband ($UBVRI$) survey reaching limiting magnitudes
$m_{AB}\sim25$~mag and covering three distinct stripes (DEEP1, DEEP2
and DEEP3) of the sky; second, the contiguous coverage in infrared of
the same regions, the focus of the present paper. The infrared
coverage also has two parts: regions covered in \k\ down to a proposed
5$\sigma$ limiting magnitude of 21.3 (in the AB system) as measured in
a 2\arcsec aperture, and a smaller area covered down to $J_{AB}=$ 23.4
and \k$_{AB}=$ 22.7~mag. As explained in Paper~III the three regions
were selected for the following reasons: 1) DEEP1 because it overlaps
with a deep radio survey; 2) DEEP2 because it includes the CDF-S field
(DEEP2c); and 3) DEEP3 due to its convenient characteristics and
location in the northern galactic cap. The optical observations were
carried out using the wide-field imager (WFI), mounted on the ESO/MPG
2.2m telescope at La~Silla covering about $0.25$ square degrees per
pointing. Each stripe consists of four adjacent pointings (denoted by
a-d in decreasing order of right ascension), yielding an area of one
square degree. Further details about DPS can be found in the EIS web
pages\footnote{http://www.eso.org/science/eis}.

The primary goal of the DPS survey is to produce a data set from which
statistical samples of galaxies can be drawn to study the large scale
structures at high redshift. However, these data are also valuable for
many other areas of investigation, in particular for the
cross-identification with X-ray sources detected by Chandra and XMM,
as a photometric reference for the ISAAC mosaic built by the GOODS
survey \citep{dickinson03,giavalisco04}, and combined with the optical
data described in Paper~III to improve photometric redshifts and to
search for rare high-redshift quasars and galaxies \citep[e.g.,
][]{kong05}.

The purpose of the present paper is to describe and present the data
from new infrared observations carried out as part of DPS; the optical
data are presented in a separate paper of this series (Paper~III). The
shallow \k\ survey is one of the largest currently available, and the
deep \j\k\ is the largest to similar depth \citep[see][for a
review]{elston05}. The paper is organized as follows:
Sect.~\ref{sec:observations} reviews the overall observing strategy
and describes the observations. Sect.~\ref{sec:reductions} briefly
describes the techniques employed to process the images, and to
astrometrically and photometrically calibrate them. This section also
includes a discussion of the assessment made of the quality of the
images and of the photometric calibration. Sect.~\ref{sec:products}
presents the final products which include the stacked images and the
single-passband catalogs extracted from them. Even though the
observations are planned as mosaics, the creation of image mosaics and
associated catalogs are beyond the scope of the present paper. Also
left for future papers are the production and analysis of
optical/infrared color catalogs. While the goal of this paper is not
to interpret the data, Sect.~\ref{sec:discussion} presents the results
of comparisons between the present observations and those of other
authors. This is done to assess the quality of the astrometry and
photometry of the data set. A brief summary is presented in
Sect.~\ref{sec:summary}.

\section{Observations}
\label{sec:observations}

\begin{table*}
\caption{The field centers and number of pointings in the shallow- and
deep-strategy mosaics.}
\label{tab:obs}
\center
\begin{tabular}{lcccccc}
\hline\hline
Field & Ra (J2000) & Dec (J2000) &$K_s$-shallow & $J$ & $K_s$-deep \\
\hline
Deep1a  &  22:55:00.0 & -40:13:00 & 49 & 16  & 16 \\
Deep1b  &  22:52:07.1 & -40:13:00 & 49 & 16  & 16 \\
Deep2a  &  03:37:27.5 & -27:48:46 & 29 & $-$  & $-$   \\
Deep2b  &  03:34:58.2 & -27:48:46 & 49 & 16  & 16  \\
Deep2c  &  03:32:29.0 & -27:48:46 & 49 & 12  & 12   \\
Deep3a  &  11:24:50.0 & -21:42:00 & 36 & 16  & 16   \\
Deep3b  &  11:22:27.9 & -21:42:00 & 36 & 16  & 16   \\
\hline
\end{tabular}
\end{table*}

The near-infrared part of the DPS was carried out using the SOFI
camera \citep{moorwood98} mounted on the New Technology Telescope
(NTT) at La~Silla. SOFI is equipped with a Rockwell 1024$^2$ detector
that, when used together with its large field objective, provides
images with a pixel scale of $0\farcs29$, and a field of view of about
$4\farcm9\times4\farcm9$. The infrared pointings were planned to
produce a contiguous coverage of the regions of interest with
sufficient overlap between adjacent pointings to enable the
construction of an image mosaic. The size of the overlaps was chosen
to avoid a significant decrease in the effective exposure time due to
the jitter pattern used, and to enable a common photometric zero-point
to be established. As mentioned earlier the main objective of the
survey was to cover each WFI pointing with shallow (720 sec) \k\
observations, complemented by 3600~sec exposures in $J$ and longer
(4800 sec) exposures in \k\ over a $4\times4$ ($\sim$
400~square~arcmin) mosaic of the central region.

The data presented in this paper were obtained as part of the ESO
Large Programmes: 164.O-0561 and 169.A-725, covering a total of 8~ESO
periods (64--73). Data for this survey were taken in
12 visitor mode runs, comprising 81 nights in the period from January
31, 2000 to September 8, 2004. The observations include 14567 science
exposures, out of a total of over 27000 frames, corresponding to a
total of about 243 hours on-target. A full log of the observations can
be found in the EIS web pages.

The observations were carried out in groups of exposures, referred to
as observation blocks (OBs), adopting the commonly used jitter
strategy as implemented in the ESO standard observing template
(AutoJitter) for the SOFI instrument. Using this template, offsets are
generated randomly within a square box of a specified side length,
chosen originally to be 40\arcsec, approximately 15\% of the SOFI
detector field of view, which was later increased to 70\arcsec
(starting November 2000) to improve the background estimate. A total
of 733 science OBs, ranging from 12 to 60 exposures, with individual
exposures from 10 to 60 seconds each, and 1169 photometric standard
OBs typically consisting of 5 exposures (for a total of 100 sec) have
been executed as part of this program.

The field centers and final status of the observations are presented
in Table~\ref{tab:obs}.  The table lists in Col.~1 field name, in
Col.~2 and 3 the position of the center of the field in J2000, and in
Cols.~4--6 the number of pointings covered for each
strategy. Typically the distance between adjacent pointings is
$4\farcm5$ and the effective area of each pointing covers about
20-25~square arcmin. The numbers listed in the table represent the
number of observed pointings regardless of their exposure
times. Completeness maps for each combination of strategy and field,
showing the ratio between actual and planned integration times, can be
found on the EIS web pages. Note that for the Deep1 fields, a number
of pointings reach only 85\% of the planned integration time for the
\k\ deep-strategy. In the case of Deep2c, only 12 pointings are listed
in the table because the remaining four are part of the EIS-DEEP
survey (see Paper~II). In keeping track of the layout of the mosaics,
the case of Deep2c is somewhat unique as the larger mosaic was built
from a smaller, pre-existing one.

From the table one finds that there are 297 pointings contributing to
the \k\ shallow strategy, corresponding to $\sim$~2.1~square~degrees,
neglecting overlaps between pointings. In addition, there are 92
pointings in both \j\ and \k\ contributing to the deep strategy. This
corresponds to $\sim$~0.63~square~degrees.  Note that in the case of
the \k\ observations the deep pointings have been included in the area
estimate of both shallow and deep strategies.

In Sect.~\ref{sec:products} the final tally of the actual coverage
achieved by the survey will be presented after taking into account the
quality of the reduced data and the final images produced by the
system.

\section{Data reduction}
\label{sec:reductions}

The data reduction was carried out using the unsupervised mode of the
EIS Data Reduction System \citep{dacosta04}, to reduce nightly images,
to determine photometric solutions on a nightly basis, and to produce
final stacked images and science-grade catalogs. A full description of
the methods used in processing the images can be found in Paper~I,
and an extensive discussion on the validation of the software in
handling infrared data is presented in Paper~II. The interested reader
is referred to these papers for more details.

The astrometric calibration of the infrared frames was performed using
the USNO-B catalog as reference following the method developed by
\cite{djamdji93}. The method is based on a multi-resolution
decomposition of images using wavelet transforms.  This package has
proven to be efficient and robust for pipeline reductions. The
estimated internal accuracy of this technique is about half a pixel
($\sim0\farcs15$) and is limited by the internal accuracy of the
reference catalog.

The photometric calibration for each night is based on observations of
standard stars taken from \cite{persson98}. Typically two to seven
stars were observed over a range of airmass. For all nights
independent photometric solutions were obtained using the photometric
pipeline of the EIS data reduction system. Depending on the airmass
and color coverage, linear fits with one to three parameters are
attempted. If no standard star observations are available a default
solution is assigned based on the solutions present in the
database. The solutions typically have errors of $\lesssim~0.03$~mag
in both passbands and negligible color terms.

\begin{table}
\caption{Type of photometric solutions available for each mosaic.}
\label{tab:ppsolutions}
\center
\begin{tabular}{lcccccc}
\hline\hline
Field  & Passband & Default & 1-par & 2-par & 3-par & total \\
\hline
Deep1a  &   $J$  & 0 &  6 & 0 & 1 & 7  \\
Deep1b  &   $J$  & 2  & 4 & 0 & 0 & 6 \\
Deep2b  &   $J$  & 1 &  0  &  0 & 7 & 8 \\
Deep2c  &   $J$  & 0  &  0 & 4 & 4 & 8 \\
Deep3a  &   $J$  & 0 &  2 & 4 & 4   & 10 \\
Deep3b  &   $J$  & 0 &  1 & 4 & 1 & 6 \\
\hline
Shallow\\
\hline
Deep1a  & $K_s$    & 0  &  1  & 2  & 0  & 3  \\
Deep1b  & $K_s$    & 1  &  0  & 3  & 1  & 5  \\
Deep2a  & $K_s$    & 2  & 1   & 2  & 1  & 6 \\
Deep2b  & $K_s$    & 1  & 1   & 4  & 6   &  12*\\
Deep2c  & $K_s$    & 1  & 1   & 6  & 6   &  14*\\
Deep3a  & $K_s$    & 2  & 4   & 5  & 3   &  14*\\
Deep3b  & $K_s$    & 1  & 4   & 4  & 1   &  10*\\
\hline
Deep\\
\hline
Deep1a  & $K_s$    & 1  &  5  & 2  & 2   & 10  \\
Deep1b  & $K_s$    & 2  &  1  & 4  & 3   & 10  \\
Deep2b  & $K_s$    & 1  & 1   & 4  & 6   & 12  \\
Deep2c  & $K_s$    & 1  & 0   & 3  & 6   &  10 \\
Deep3a  & $K_s$    & 2  &  3  & 4  & 3   &  12 \\
Deep3b  & $K_s$    & 1  &  3  & 4  & 1   &  9 \\
\hline
\multicolumn{7}{l}{* includes Deep}\\
\end{tabular}
\end{table}

Table~\ref{tab:ppsolutions} summarizes the type of photometric
solution for each mosaic. The table gives in Col.~1 the field name, in
Col.~2 the passband, in Col.~3 the number of default solutions, in
general corresponding to non-photometric nights, and in Cols. 4--6 the
number of nights with photometric solutions derived from 1- to 3-
parameter fits and in Col.~7 the total number of nights used to
observe pointings in a given mosaic. In most of the mosaics solutions
with at least 2 parameters are available thereby assuring an adequate
calibration of at least some frames in each mosaic. The only exception
is Deep1b in \j, where only 1-parameter solutions were possible,
leading to a less accurate calibration of these data.  In only one of
the remaining cases (Deep1a \k\ shallow) no 3-parameter solution is
available.

\begin{table}
\caption{Average photometric solutions} 
\label{tab:zp} 
\center
\begin{tabular}{cccc}
\hline\hline
Passband & Zp & k & color\\
\hline
\j&23.12$\pm$0.09&0.08$\pm$0.04&-0.02$\pm$0.05 \\
\k&22.41$\pm$0.08&0.08$\pm$0.06&-0.01$\pm$0.24 \\
\hline
\end{tabular}
\end{table}

\begin{table*}
\caption{Summary of reductions.}
\label{tab:red}
\center
\begin{tabular}{lccccccc}
\hline\hline
Field & Passband & \# Pointings & T (ksec) & Nights & Exposures & \#
Images \\
\hline
Deep1a  & $J$  & 16 & 60.4  & 7 & 1006  & 40 \\
Deep1b & $J$   & 13 & 53.3  & 6 & 888   & 31 \\
Deep2b  & $J$  & 16 & 60.4  & 8 & 1007  & 41  \\
Deep2c  & $J$  & 12 & 45.4  & 8 & 756   & 14  \\
Deep3a  & $J$  & 16 & 63.4  &10 & 1056  & 22 \\
Deep3b  & $J$  & 16 & 57.5  & 6 & 958  & 32 \\
\hline
Shallow\\
\hline
Deep1a  & $K_s$  & 36  & 26.8  & 3  &  447 & 39\\
Deep1b  & $K_s$  & 30  & 20.7  & 5  &  345 & 35\\
Deep2a  & $K_s$  & 29  & 20.0  & 6  &  334 & 29\\
Deep2b  & $K_s$  & 36  & 89.6  & 12 & 1494 & 78*\\
Deep2c  & $K_s$  & 49  & 82.5  & 14 & 1375 & 66*\\
Deep3a  & $K_s$  & 36  & 89.8  & 14 & 1498 & 54*\\
Deep3b  & $K_s$  & 35  & 94.3  & 10 & 1572 & 56*\\
\hline
Deep\\
\hline
Deep1a  & $K_s$  & 14  & 51.6  & 10 & 860 & 32\\
Deep1b  & $K_s$  & 14  & 52.5  & 10 & 876 & 32\\
Deep2b  & $K_s$  & 16  & 75.7  & 12 & 1262 & 58\\
Deep2c  & $K_s$  & 12  & 57.3  & 10 & 956  & 28\\
Deep3a  & $K_s$  & 16  & 75.9  & 12 & 1266 & 34\\
Deep3b  & $K_s$  & 16  & 80.9  & 9  & 1349 & 37\\
\hline
\multicolumn{7}{l}{* includes Deep}\\
\end{tabular}
\end{table*}

Table~\ref{tab:zp} summarizes the average photometric solutions
determined for the entire period of observations. For each passband
the table lists in Col.~1 the passband, in Col.~2 the mean and
standard deviation of the measured zero points, in Col.~3 the mean and
standard deviation of the determined extinction coefficients, in
Col.~4 the mean and standard deviation of the measured color
terms. The listed standard deviations are a good indication of the
uncertainty of the estimated zero points.

The results of the science reduction are summarized in
Table~\ref{tab:red}.  This table lists in Col.~1 the field name,
in Col.~2 the passband, in Col.~3 the number of pointings observed, in
Col.~ 4 the total on-source time in ksec, in Col.~5 the number of
nights in which the observations were carried out, in Col.~6 the
number of raw exposures involved, and in Col.~7 the number
of reduced images available. A total of 609 reduced images were
produced from the 14567 science exposures corresponding to a total
integration time of 874 ksecs ($\sim$ 243 hours).

All the reduced images were inspected and graded and in
Table~\ref{tab:qc} the results of the visual inspection are summarized
grouping by mosaic. The table lists in Col.~1 the name of the field,
in Col.~2 the passband, and in Cols.~3--6 the grades ranging from A
(best) to D (worst).  These grades reflect the overall cosmetic
quality of the images including the background and the presence of
other features that may affect the source extraction.  In addition to
the grade a subjective comment is normally associated with the image,
to describe any abnormality or feature relevant to a user.  Normally
the comments are related to poor background estimation or remaining
imprints of the fringing in particular in the \k -band images.
Typically, the $J$-band images are of excellent quality with over 80\%
being graded A. The worse grades are usually due to the presence of
bright vertical lines resulting from cross-talk effects associated with
the presence of relatively bright stars in or near the field. In the
\k -band $\sim65\%$ are graded A and an additional $\sim20\%$ are
graded B. These grades, on average lower, reflect in particular that
residual fringing is more common in this band.

\begin{table}
\caption{Quality assessment of reduced images. }
\label{tab:qc}
\center
\begin{tabular}{lccccccc}
\hline\hline
Field & Passband &  A  &  B   & C  & D \\
\hline
Deep1a  & $J$  &  34 & 6  &  0 & 0 \\
Deep1b  & $J$  &  18 & 12  &  0 & 1 \\
Deep2b  & $J$  &  35 & 6  &  0 & 0 \\
Deep2c  & $J$  &  12  & 1  &  0 & 1 \\
Deep3a  & $J$  &  19  & 2  &  1 & 0 \\
Deep3b  & $J$  &  28  & 4  &  0 & 0 \\
\hline
Shallow\\
\hline
Deep1a  & $K_s$  & 16  & 18  &  4 & 1 \\
Deep1b  & $K_s$  & 10  & 14  &  6 & 5 \\
Deep2a  & $K_s$  & 28  & 4   & 5  & 0 \\
Deep2b  & $K_s$  & 60  & 11  &  6 & 1 *\\
Deep2c  & $K_s$  & 55  & 6   & 3  & 2 *\\
Deep3a  & $K_s$  & 42  & 10  &  2 & 0 *\\
Deep3b  & $K_s$  & 38  & 12  &  5 & 1 *\\
\hline
Deep\\
\hline
Deep1a  & $K_s$  & 6   & 14 & 12 & 0 \\
Deep1b  & $K_s$  & 17  & 10 &  3 & 2 \\
Deep2b  & $K_s$  & 41  & 10 &  6 & 1 \\
Deep2c  & $K_s$  & 26  & 1  &  1 & 0 \\
Deep3a  & $K_s$  & 27  & 7  &  0 & 0 \\
Deep3b  & $K_s$  & 25  & 8  &  3 & 1 \\
\hline
\multicolumn{6}{l}{* includes Deep}\\
\end{tabular}
\end{table}

\section {Final products}
\label{sec:products}

\subsection{Images}

Final stacked images were produced from the nightly reduced images,
described in the previous section, properly grouped in \emph{ Stacking
Blocks} (SBs) as explained in detail in Paper~I. The 178 and 417
reduced images with grades better than D in the \j- and \k-bands, were
converted into 89 and 272 stacked (co-added) images. The process of
producing the final stacks takes into account differences in zero
point that exist for individual images by correcting all frames to the
zero airmass flux level. Non-photometric frames are scaled to the zero
airmass flux level of the photometric frames, thereby assuring the
photometric quality of the produced images.

\begin{table}
\caption{Quality assessment of stacked images. }
\label{tab:qc_stacks}
\center
\begin{tabular}{lccccccc}
\hline\hline
Field & Passband &  A  &  B   & C  & D \\
\hline
Deep1a  & $J$  &  14 & 2  &  0 & 0 \\
Deep1b  & $J$  &  11 & 2  &  0 & 0 \\
Deep2b  & $J$  &  16 & 0  &  0 & 0 \\
Deep2c  & $J$  &  11 & 1  &  0 & 0 \\
Deep3a  & $J$  &  14 & 2  &  0 & 0 \\
Deep3b  & $J$  &  14 & 2  &  0 & 0 \\
\hline
Shallow\\
\hline
Deep1a  & $K_s$  & 17  & 10  & 8  & 0 \\
Deep1b  & $K_s$  & 10  & 11  & 5  & 1 \\
Deep2a  & $K_s$  & 27  & 1   & 1  & 0 \\
Deep2b  & $K_s$  & 33  & 3   & 0  & 0 *\\
Deep2c  & $K_s$  & 40  & 3   & 3  & 0 *\\
Deep3a  & $K_s$  & 23  & 8   & 3  & 2 *\\
Deep3b  & $K_s$  & 26  & 4   & 3  & 2 *\\
\hline
Deep\\
\hline
Deep1a  & $K_s$  & 2   & 5  &  6 & 1 \\
Deep1b  & $K_s$  & 10  & 4  &  0 & 0 \\
Deep2b  & $K_s$  & 15  & 1  &  0 & 0 \\
Deep2c  & $K_s$  & 11  & 0  &  1 & 0 \\
Deep3a  & $K_s$  & 8   & 6  &  2 & 0 \\
Deep3b  & $K_s$  & 12  & 2  &  2 & 0 \\
\hline
\multicolumn{6}{l}{* includes Deep}\\
\end{tabular}
\end{table}

As before all images were visually inspected and graded. Out of the
361 stacked images covering the 7 DPS-NIR mosaics, 268 were graded A,
56 B, 31 C, and 6 D. In Table~\ref{tab:qc_stacks} the grade
distribution for each mosaic is given. The table lists in Col.~1 the
field name, in Col.~2 the passband, and in Cols.~3--6 the number of
images in each mosaic with grade A--D, respectively. As in the case of
the reduced images the comments made for the stacks are normally
related to residual fringing and poor background subtraction, both of
which are most frequent in the \k -band. In \j -band the most frequent
problem is cross-talk features.

\begin{table*}
\center
\caption{Stacked image attributes and completeness for each mosaic
only considering pointings with grade better then D.}
\label{tab:attributes}
\begin{tabular}{llccrcccc}
\hline
\hline
Field & Passband & \# Pointings & PSF FWHM & PSF rms
& $m_{lim}$ (Vega) & Completeness \\
\hline
Deep1a & \j & 16&0.676$\pm$0.094&0.086$\pm$0.025&22.17$\pm$ 0.23&104\%$\pm$ 27\%\\
Deep1b & \j & 13&0.935$\pm$0.238&0.073$\pm$0.034&22.14$\pm$ 0.56& 86\%$\pm$ 38\%\\
Deep2b & \j & 16&0.791$\pm$0.191&0.110$\pm$0.041&22.06$\pm$ 0.16&103\%$\pm$ 12\%\\
Deep2c & \j & 12&0.837$\pm$0.311&0.090$\pm$0.034&22.01$\pm$ 0.17& 96\%$\pm$ 11\%\\
Deep3a & \j & 16&0.751$\pm$0.141&0.090$\pm$0.033&22.02$\pm$ 0.49&110\%$\pm$ 40\%\\
Deep3b & \j & 16&0.712$\pm$0.066&0.076$\pm$0.016&22.05$\pm$ 0.17& 99\%$\pm$  0\%\\
\hline
Shallow\\
\hline
Deep1a & \k & 35&1.275$\pm$0.466&0.053$\pm$0.021&19.57$\pm$ 0.16&103\%$\pm$ 25\%\\
Deep1b & \k & 26&0.890$\pm$0.198&0.065$\pm$0.040&19.38$\pm$ 0.29& 97\%$\pm$ 30\%\\
Deep2a & \k & 29&0.749$\pm$0.274&0.066$\pm$0.023&19.43$\pm$ 0.20& 95\%$\pm$ 11\%\\
Deep2b & \k & 36&0.715$\pm$0.181&0.064$\pm$0.025&19.47$\pm$ 0.16& 96\%$\pm$  8\%*\\
Deep2c & \k & 46&1.387$\pm$0.602&0.057$\pm$0.023&19.62$\pm$ 0.14& 96\%$\pm$ 21\%*\\
Deep3a & \k & 34&0.737$\pm$0.092&0.059$\pm$0.011&19.39$\pm$ 0.08&100\%$\pm$  0\%*\\
Deep3b & \k & 33&0.688$\pm$0.097&0.052$\pm$0.009&19.30$\pm$ 0.09&100\%$\pm$  0\%*\\
\hline
Deep\\
\hline
Deep1a & \k & 13&0.712$\pm$0.090&0.095$\pm$0.026&20.07$\pm$ 0.23& 77\%$\pm$ 16\%\\
Deep1b & \k & 14&0.911$\pm$0.209&0.089$\pm$0.026&20.24$\pm$ 0.31& 77\%$\pm$ 21\%\\
Deep2b & \k & 16&0.842$\pm$0.151&0.144$\pm$0.039&20.44$\pm$ 0.17& 97\%$\pm$  5\%\\
Deep2c & \k & 12&0.671$\pm$0.080&0.133$\pm$0.043&20.30$\pm$ 0.20& 99\%$\pm$ 23\%\\
Deep3a & \k & 16&0.712$\pm$0.159&0.102$\pm$0.027&20.34$\pm$ 0.14& 98\%$\pm$  9\%\\
Deep3b & \k & 16&0.679$\pm$0.080&0.104$\pm$0.020&20.26$\pm$ 0.12&101\%$\pm$ 18\%\\
\hline
\multicolumn{6}{l}{* includes Deep}\\
\end{tabular}
\end{table*}

The main attributes of the stacks produced for each mosaic are
summarized in Table~\ref{tab:attributes}. The table gives in Col.~1
the field name, in Col.~2 the passband, in Col.~3 the number of
pointings available with grade better than D, in Cols.~4 and 5 the
average and standard deviation of the FWHM in arcseconds and anisotropy
(PSF rms) of the point-spread function measured in the final
stacks, in Col.~6 the average and standard deviation of the limiting
magnitude, $m_{lim}$, estimated for the final image stack for a
2\arcsec\ aperture, $5\sigma$ detection limit in the Vega system, in
Col.~7 the average and standard deviation of the completeness
expressed as the fraction (in percentage) of observing time relative
to that originally planned.

The overall image quality is high with the average PSF having a FWHM
of $\lesssim1''$ and an anisotropy typically less than 10\%. Two of
the shallow mosaics (Deep1a and Deep2c) show a significantly larger
FWHM. This is because in these mosaics there are a few frames taken in
poor seeing conditions, and does not reflect the quality of these
mosaics as a whole.  The overall median limiting magnitudes in the
Vega system are 22.16~mag ($J_{AB}=23.06$) in \j-band;
$K_s$=19.57~mag ($K_{sAB}=21.41$) for the shallow strategy;
$K_s$=20.32~mag ($K_{sAB}=22.16$) for the deep strategy.  This means
that the proposed depth of \k$_{AB}=21.3$~mag for the \k-band shallow
has on average been reached, while the deep \j\k\ surveys are about
0.3-0.5 magnitudes shallower than originally planned ($J_{AB}=23.4$
and $K_{sAB}=22.7$~mag).

The completeness of the survey is measured not only in terms of depth
but also in areal coverage. The latter can be assessed from
Figs.~\ref{fig:deep1_coverage}-\ref{fig:deep3_coverage}, in the
appendix. These figures show the final coverage for each region and
strategy, after discarding images with grade D. Note that in the case
of Deep1 the \k\ deep observations were done independently of those
conducted for \k-shallow.  As can be seen the areal completeness and
connectivity of the mosaics are high, with the exception of the
shallow observations of Deep1a and b, which are 70\% and 53\% areal
complete, respectively.

\subsection{Source lists}
\label{sec:catalogs}

\begin{figure*}
\resizebox{\textwidth}{!}{\includegraphics{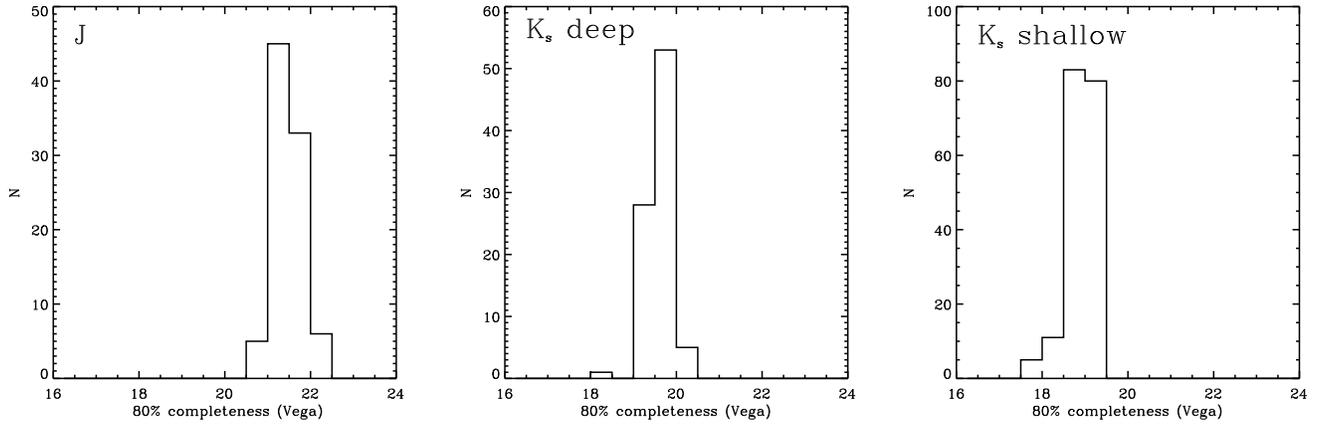}}
\caption{Distribution of the 80\% completeness limiting Vega-magnitudes
 for each of the three observing strategies as indicated in each
 panel.}
\label{fig:cat_attributes}
\end{figure*}

\begin{table*}
\center
\caption{Summary of the properties of the catalogs extracted for each mosaic and band.}
\label{tab:catalogs}
\begin{tabular}{llrccc}
\hline
\hline
Field & Passband & \#Catalogs & Eff. area & \#Objects & 80\% compl.\\
\hline
Deep1a & \j &       16 &  29.4$\pm$  1.0 &          908$\pm$         110 &  21.50$\pm$  0.23\\
Deep1b & \j &       13 &  28.7$\pm$  1.8 &          665$\pm$         145 &  21.57$\pm$  0.51\\
Deep2b & \j &       16 &  29.4$\pm$  1.0 &          828$\pm$          90 &  21.48$\pm$  0.23\\
Deep2c & \j &       12 &  28.8$\pm$  2.1 &          811$\pm$         208 &  21.41$\pm$  0.25\\
Deep3a & \j &       16 &  27.2$\pm$  2.2 &          934$\pm$         178 &  21.52$\pm$  0.38\\
Deep3b & \j &       16 &  29.4$\pm$  0.7 &          907$\pm$          80 &  21.44$\pm$  0.25\\
\hline
Shallow\\
\hline
Deep1a & \k &       35 &  26.0$\pm$  1.1 &          361$\pm$         161 &  18.94$\pm$  0.38\\
Deep1b & \k &       26 &  24.6$\pm$  1.7 &          484$\pm$         112 &  18.95$\pm$  0.24\\
Deep2a & \k &       29 &  26.1$\pm$  1.6 &          440$\pm$          91 &  19.05$\pm$  0.22\\
Deep2b & \k &       20 &  26.1$\pm$  1.2 &          479$\pm$          78 &  19.05$\pm$  0.17\\
Deep2c & \k &       34 &  25.7$\pm$  1.6 &          318$\pm$         188 &  18.79$\pm$  0.47\\
Deep3a & \k &       18 &  26.3$\pm$  0.5 &          593$\pm$          98 &  19.01$\pm$  0.09\\
Deep3b & \k &       17 &  26.7$\pm$  0.6 &          551$\pm$          53 &  18.91$\pm$  0.08\\
\hline
Deep\\
\hline
Deep1a & \k &       13 &  29.4$\pm$  1.1 &          996$\pm$         133 &  19.40$\pm$  0.22\\
Deep1b & \k &       14 &  28.9$\pm$  1.2 &          660$\pm$         117 &  19.57$\pm$  0.48\\
Deep2b & \k &       16 &  28.1$\pm$  0.8 &          697$\pm$          61 &  19.70$\pm$  0.21\\
Deep2c & \k &       12 &  28.6$\pm$  1.7 &          699$\pm$          80 &  19.52$\pm$  0.21\\
Deep3a & \k &       16 &  26.9$\pm$  1.9 &          804$\pm$         126 &  19.63$\pm$  0.18\\
Deep3b & \k &       16 &  29.8$\pm$  0.6 &          932$\pm$         161 &  19.63$\pm$  0.12\\
\hline
\multicolumn{5}{l}{* includes Deep}\\
\end{tabular}
\end{table*}

Catalogs were extracted from the stacked images using the EIS Data
Reduction System adopting the same procedures as described in Papers~I
and~II. A description of the catalog format is found in the appendices
of Paper~I.  These catalogs include only objects with $S/N \geq 3$ as
computed from the magnitude error. Table~\ref{tab:catalogs} summarizes
some of the main characteristics of the extracted catalogs. The table
gives in Col.~1 the field name, in Col.~2 the passband, in Col.~3 the
average and standard deviation of the effective area in square arcmin,
in Col.~4 the average and standard deviation of the number of objects,
and in Col.~5 the average and standard deviation of the 80\%
completeness limit in the Vega system. These completeness limits are
for extended objects (see Paper~I for details). Their distribution for
each strategy is illustrated in Fig.~\ref{fig:cat_attributes}. There
is a total of 89 \j\ and 87 \k\ catalogs with mean values of the 80\%
completeness magnitude of $21.49 \pm 0.31$ and $19.57 \pm 0.28$~mag in
the Vega system, with each catalog covering an area of
$\sim28.8$~square arcmin. For the shallow \k\ strategy there are 179
images with a mean 80\% completeness magnitude in the Vega system of
$18.94 \pm 0.31$~mag, with a typical coverage of $\sim25.8$~square
arcmin.

Inspection of the {\tt CLASS\_STAR} parameter as a function of magnitude
indicates that stars and galaxies can be adequately classified by
using {\tt CLASS\_STAR} with a threshold of 0.90 at magnitudes
$K_s\leq 17.0$ in both the deep and shallow data sets.  For the
$J$-band we identify stars brighter than $J=18.0$. At fainter
magnitudes we take all objects to be galaxies.  In the next section,
these catalogs are used to assess the scientific quality of the survey
data.

\section{Discussion}
\label{sec:discussion}

\subsection{Internal photometric comparison}

\begin{table}
\caption{Mean magnitude offsets and standard deviations for each mosaic}
\label{tab:int_phot_check}
\begin{tabular}{lcccc}
\hline\hline
Field & Passband & \# overlaps & Mag. offset & Std.dev.\\
\hline
Deep1a & \j & 24 & -0.02 & 0.11\\
Deep1b & \j & 15 & 0.05 & 0.11\\
Deep2b & \j & 24 & 0.03 & 0.05\\
Deep2c & \j & 12 & -0.03 & 0.08\\
Deep3a & \j & 24 & -0.02 & 0.06\\
Deep3b & \j & 24 & 0.02 & 0.04\\
\hline
Shallow\\
\hline
Deep1a & \k & 36 & 0.07 & 0.13\\
Deep1b & \k & 21 & 0.02 & 0.18\\
Deep2a & \k & 39 & 0.08 & 0.12\\
Deep2b & \k & 8  & 0.06 & 0.22\\
Deep2c & \k & 34 & 0.05 & 0.13\\
Deep3a & \k & 15 & 0.06 & 0.08\\
Deep3b & \k & 14 & 0.01 & 0.10\\
\hline
Deep\\
\hline
Deep1a & \k & 14 & 0.00 & 0.10\\
Deep1b & \k & 16 & 0.03 & 0.11\\
Deep2b & \k & 24 & 0.03 & 0.09\\
Deep2c & \k & 12 & 0.00 & 0.08\\
Deep3a & \k & 19 & 0.03 & 0.07\\
Deep3b & \k & 24 & 0.05 & 0.08\\
\hline
\end{tabular}
\end{table}

The consistency of the photometric calibration of the individual
pointings, and therefore of the mosaic, was assessed by comparing the
magnitudes of objects in common between adjacent frames.  For each
mosaic the mean and standard deviation of the magnitude difference of
all pairs of sources in common in adjacent frames were computed,
considering only objects with a magnitude error $<0.2$
($\sim5\sigma$). Ideally, one would use only stars for this, but
because the number of objects in the overlaps is small, it proved
necessary to include galaxies as well. In
Table~\ref{tab:int_phot_check} these mean offsets and standard
deviations are listed for each mosaic. The table lists in Col.~1 the
field name, in Col.~2 the passband, in Col.~3 the number of overlaps
in the mosaic, in Cols.~4 and 5 the mean and standard deviation of the
magnitude offsets. From the table one finds that on average the
offsets are small, typically $\lesssim0.05$~mag. The scatter is
typically less than 0.1~mag. In all cases the mean offset is smaller
than the standard deviation. Combined these results suggest a
homogeneous photometric calibration across the mosaics with fairly
small field-to-field variations.

In an attempt to improve the match among different zero points obtained
on different nights, correct for data taken under non-photometric
conditions, and hence minimize field-to-field variations, the method
employed by \cite{maddox90} was utilized. Taking into account the
computed corrections, one finds that, in general, the mean offsets do
not change significantly and the scatter decreases only
marginally. This result may be partly due to the large uncertainty in
each individual offset caused by the small number of sources in the
overlap regions and the necessity to include galaxies. However, it
may also indicate a good match among the original zero points.

\subsection{External comparison}

\begin{figure*}
\begin{center}
\resizebox{0.4\textwidth}{!}{\includegraphics{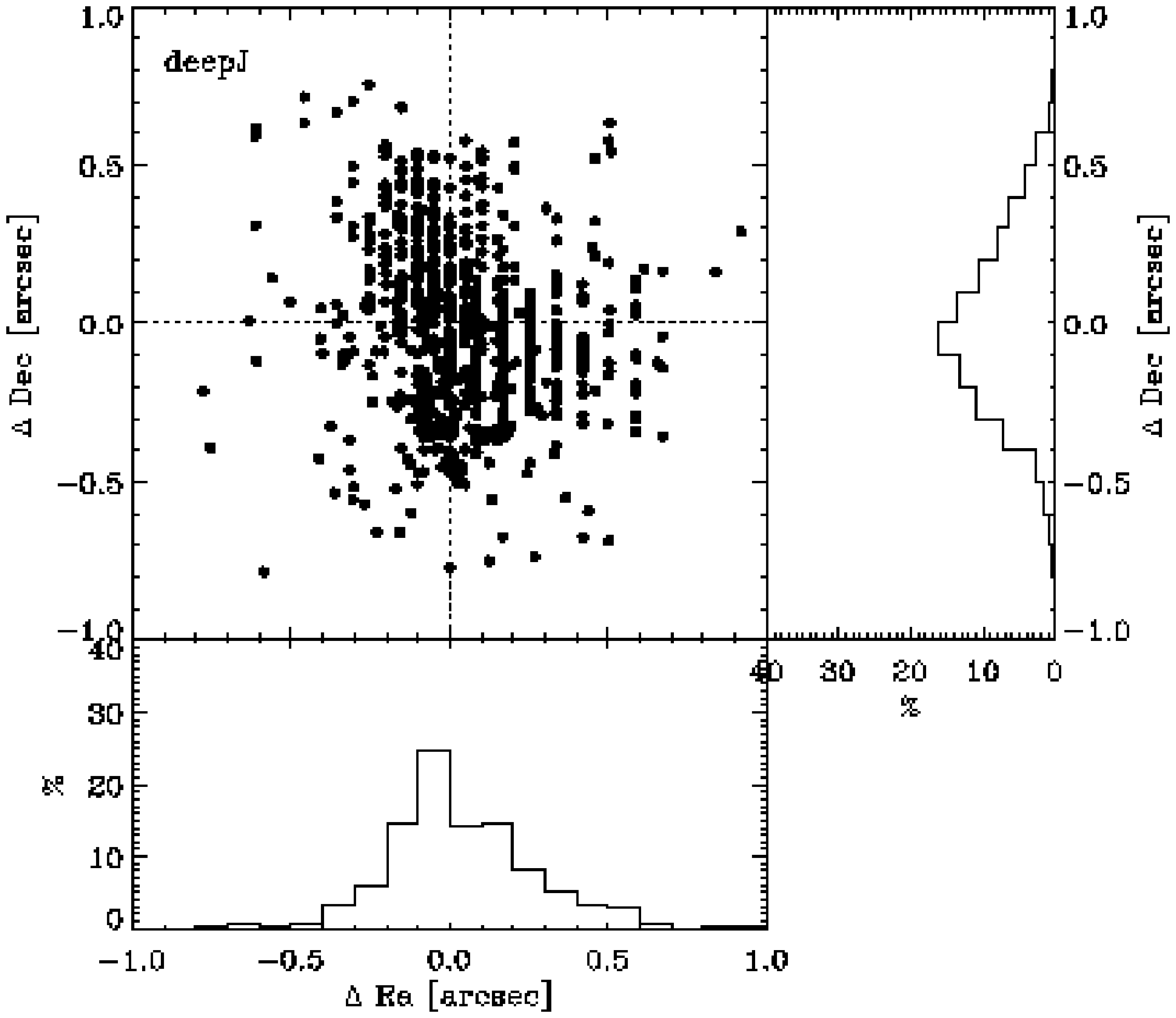}}
\resizebox{0.4\textwidth}{!}{\includegraphics{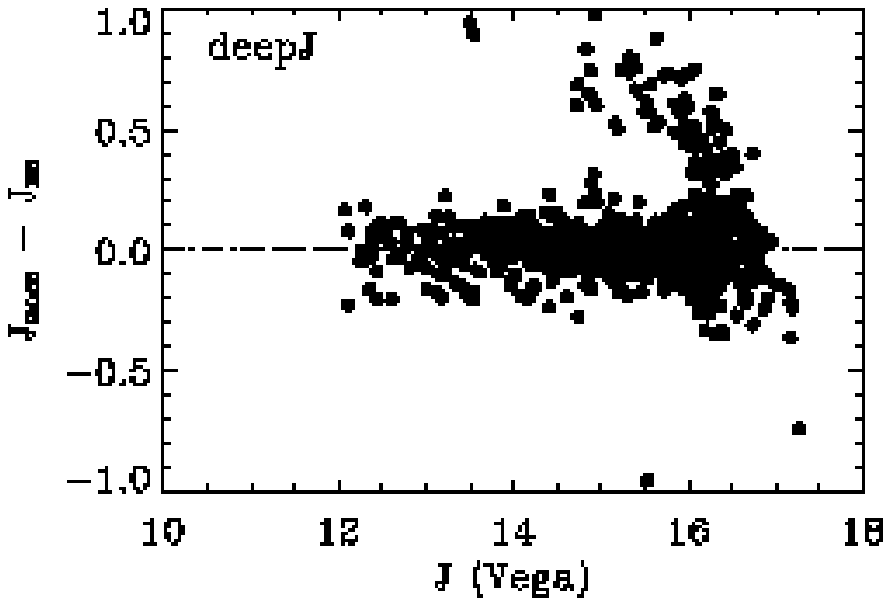}}
\resizebox{0.4\textwidth}{!}{\includegraphics{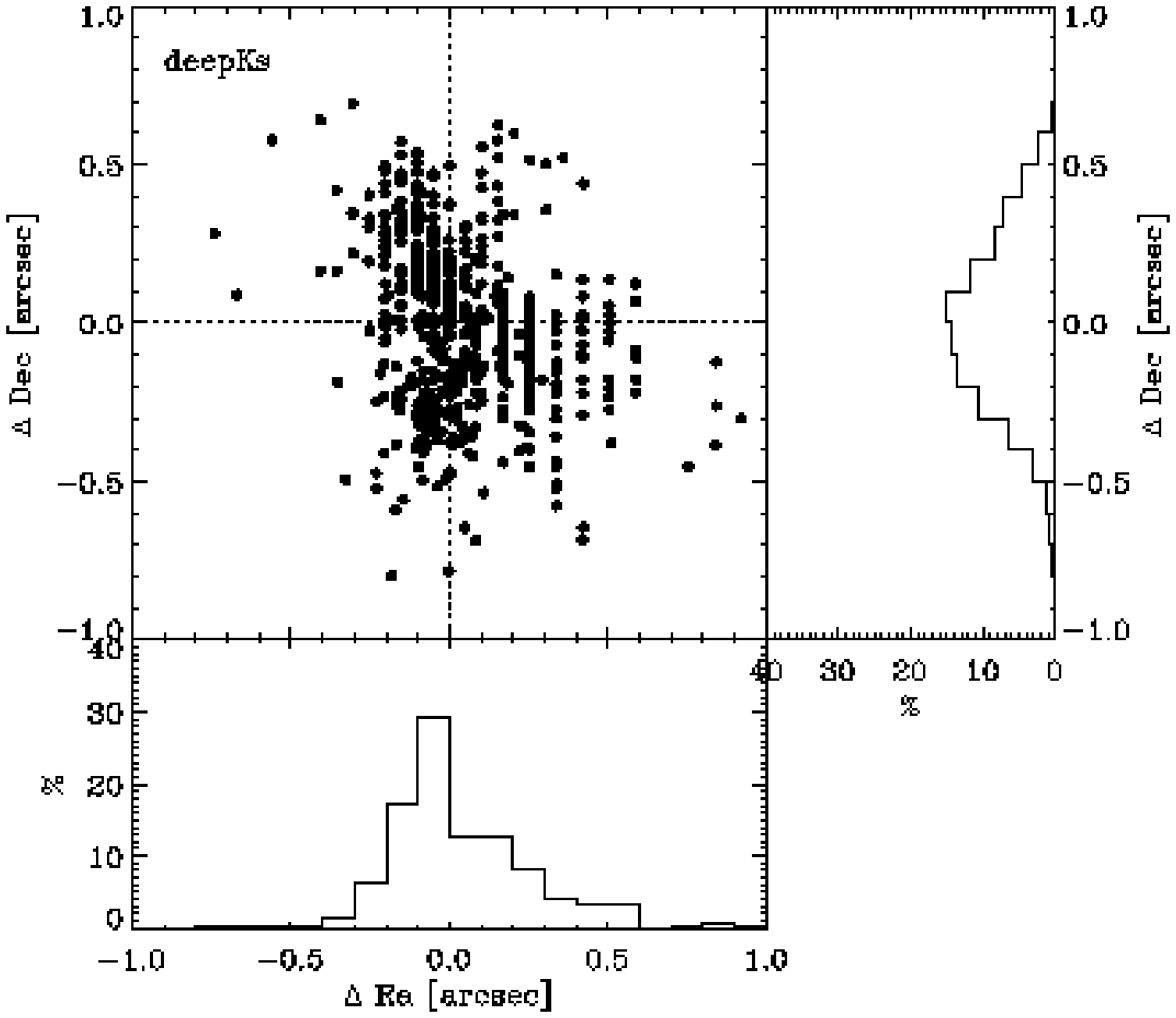}}
\resizebox{0.4\textwidth}{!}{\includegraphics{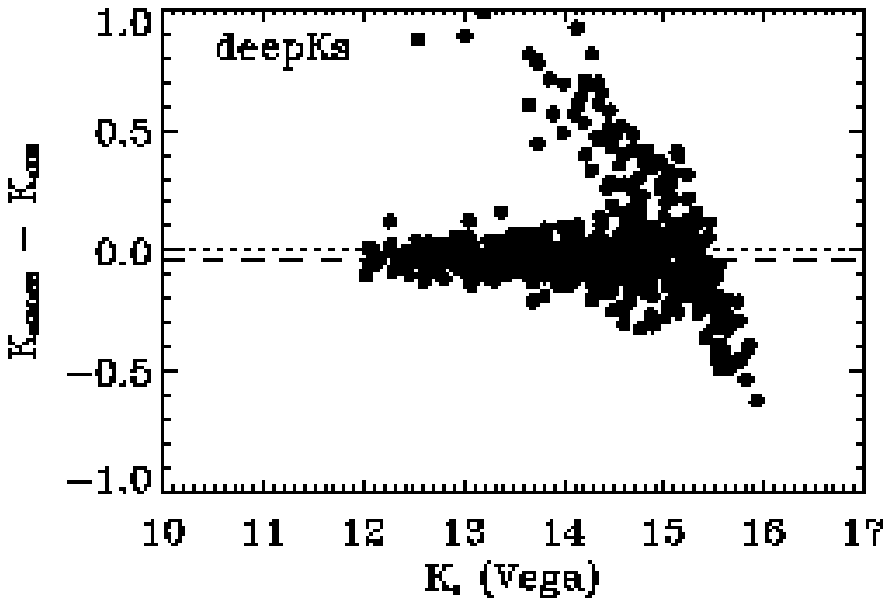}}
\resizebox{0.4\textwidth}{!}{\includegraphics{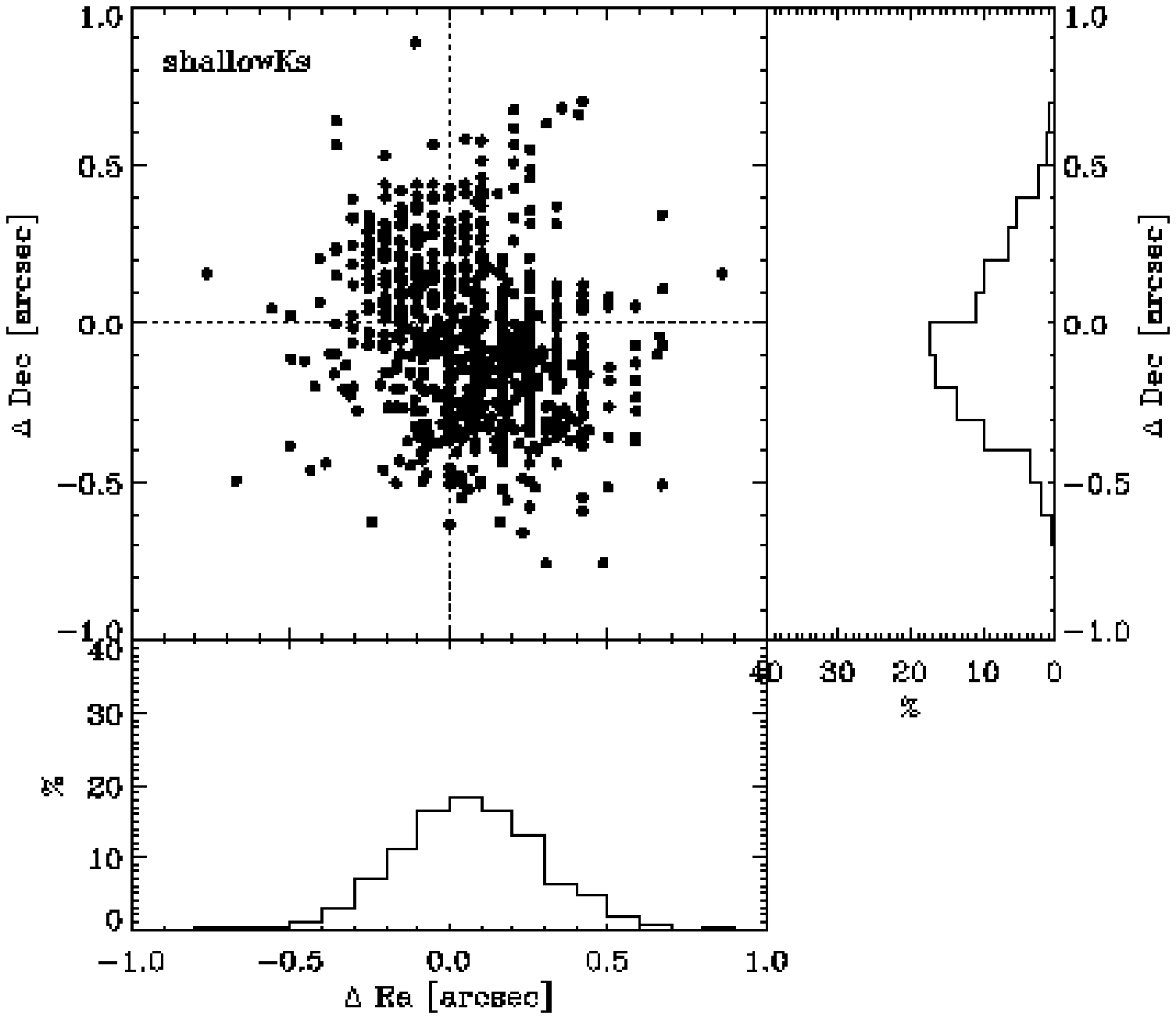}}
\resizebox{0.4\textwidth}{!}{\includegraphics{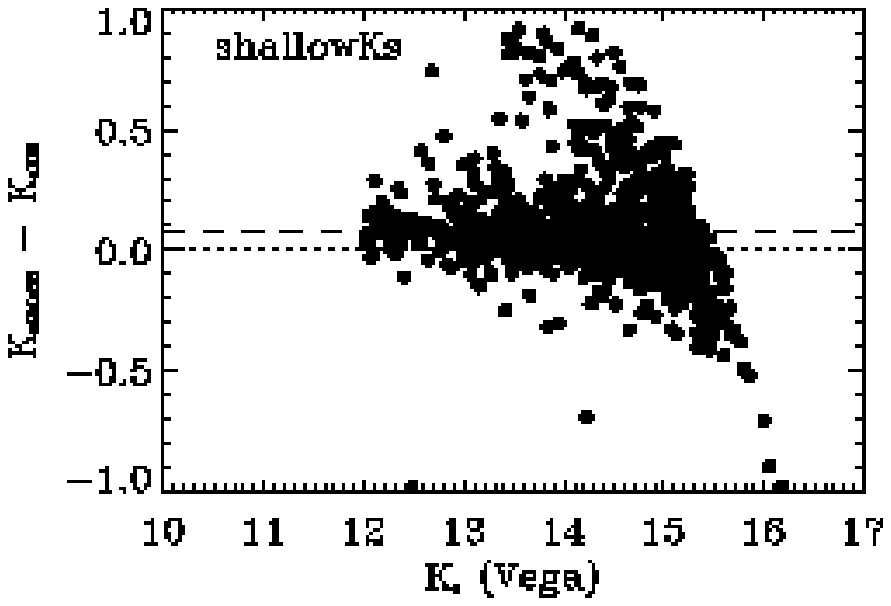}}
\end{center}
\caption{Astrometric (left column) and photometric (right column) 
comparison between objects in common between the EIS and 2MASS
surveys.  The offsets are computed as 2MASS~$-$~EIS.  The rows
correspond to each adopted strategy as indicated. Dotted lines
indicate the zero offset location, while the dashed lines in the right
panels depict the mean offset computed for the bright objects as
discussed in the text. In this plot  magnitudes are in the Vega system.}
\label{fig:comp-2mass}
\end{figure*}

In order to externally verify the survey products, the source lists
extracted from the final images as described in the previous section
were cross-correlated with the point source catalog available from the
2MASS survey \citep{skrutskie06}. This was carried out adopting a
search radius of 1\arcsec.  A total of 1056, 966, 1539 objects were
found in common for the \j, and the deep and shallow \k~-band final
images, respectively, over all regions covered by the survey.  The
differences in position and magnitude for objects in common and
considered ``good'' in both catalogs (5$\sigma$ detections and in the
case of 2MASS not contaminated by neighboring objects) were computed
and their distribution as a function of right ascension and
declination (left panels) and magnitude (right panels) are shown in
Fig.~\ref{fig:comp-2mass}. The mean offset in position is negligible
and amounts to at most $\sim0\farcs07$ for the shallow
\k~strategy. The scatter is of $0\farcs23$ and $0\farcs27$ in right
ascension and declination, respectively, for all strategies. This is
consistent with an astrometric accuracy of $\sim0\farcs19$ for the EIS
images, limited by the typical internal accuracy of the reference
catalog used.

Considering objects brighter than $J=16.0$ and $K_s=14.5$ (Vega
system) the measured magnitudes of the two data sets are in good
agreement. At fainter magnitudes the scatter increases and the effects
of the Malmquist bias introduced by the depth of the 2MASS data are
seen.  Objects with magnitudes brighter than 12~mag are saturated in
the EIS data and therefore discarded from the plots.  For the deep
strategies the systematic offset in magnitude is (discarding the
outlying objects with offsets in excess of 0.5~mag) $\sim -0.04$~mag
with a scatter of 0.07 in \k\ and $\sim0.0$ with a scatter of 0.08 in
\j , consistent with the estimated uncertainty of the overall
zero-point (Sect.~\ref{sec:reductions}). For the shallow \k\ -survey
one finds a mean offset of $\sim0.08$~mag and a scatter of
$\sim0.13$~mag. The amplitude of the scatter is remarkably small,
considering the values obtained in the previous section, and
demonstrates the uniformity of the photometric calibration of the
present data across the sky.

\subsection{Comparison of galaxy and star counts}

\begin{figure*}
\center
\resizebox{0.8\textwidth}{!}{\includegraphics{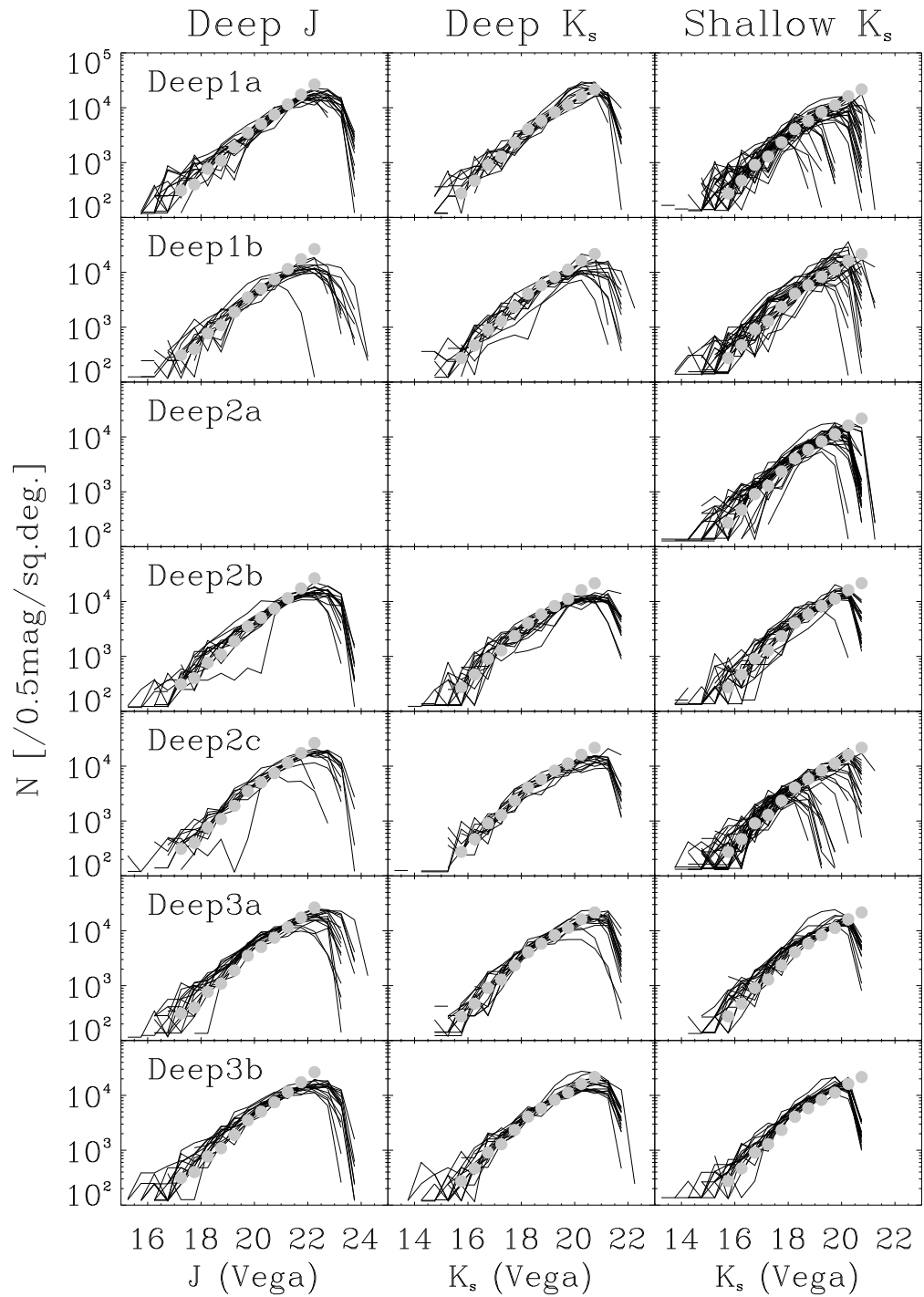}}
\caption{Galaxy counts for each individual pointing grouped by mosaic 
(lines) compared with those of \citet[ solid circles]{iovino05}.}
\label{fig:galcounts_all}
\end{figure*}

\begin{figure}
\begin{center}
\resizebox{\columnwidth}{!}{\includegraphics{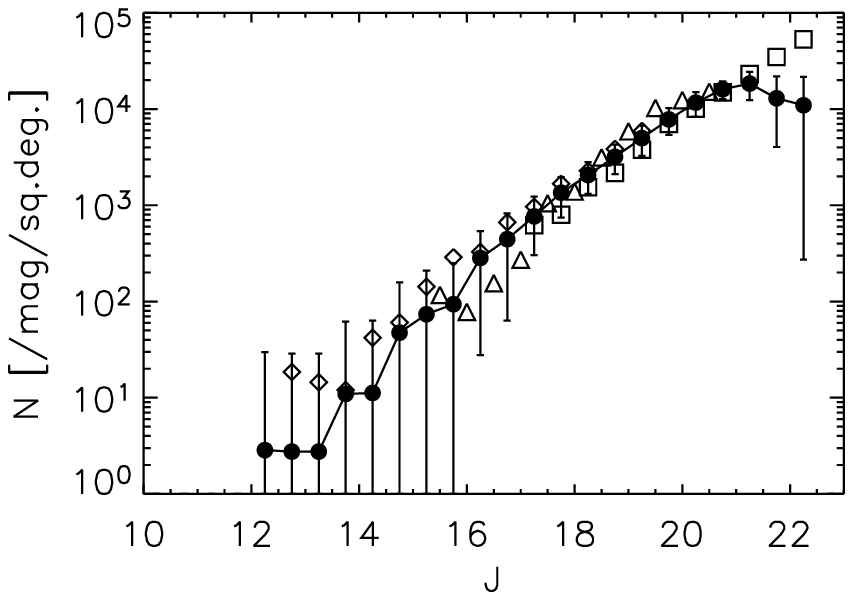}}
\resizebox{\columnwidth}{!}{\includegraphics{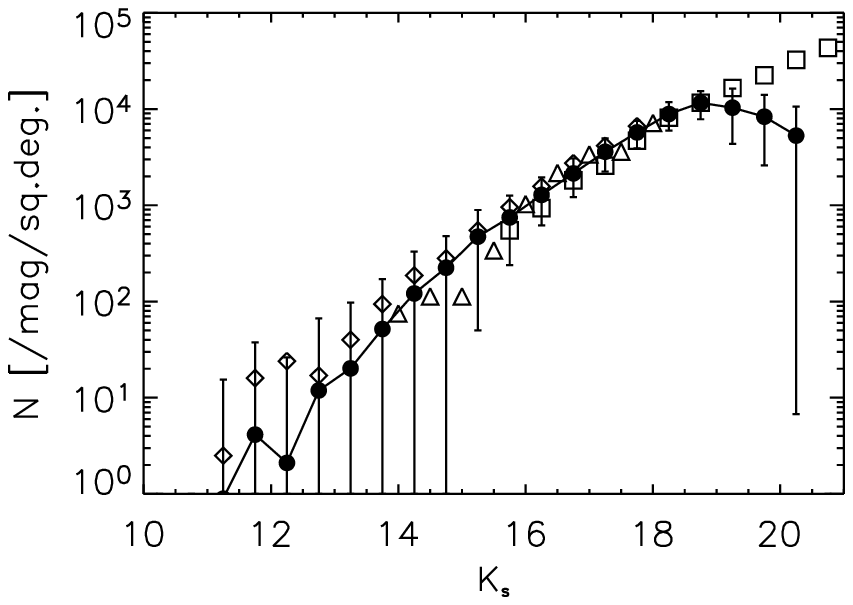}}
\end{center}
\caption{Comparison of galaxy number counts in the $J$- (top) and
$K_s$-band (bottom) between the present work (solid circles with
error bars); diamonds denote \cite{vaisanen00}, triangles
\cite{martini01} and squares \cite{iovino05}. }
\label{fig:counts_avrg}
\end{figure}

An alternative way to evaluate the data is to compare the counts of
galaxies and stars with those computed by other authors and/or
predicted by models. For galaxies this is illustrated in
Fig.~\ref{fig:galcounts_all}. The figure shows for each passband and
strategy (columns) the galaxy counts obtained for each pointing (thin
lines) grouped by mosaic (rows), as indicated in the panels of the
leftmost panel of each row. These are compared with the counts
recently reported by \citet[ solid circles]{iovino05}. Here galaxies
are 3$\sigma$ detections with CLASS\_STAR$<0.9$. From the figure one
can easily: 1) identify individual cases that depart significantly
from the mean; 2) have a clear measure of the variance introduced from
field-to-field variations due to the relatively small field of view of
SOFI; and 3) identify the variation in depth of a few pointings, such
as in the case of Deep1a \k-shallow among others. Moreover, the
relatively small scatter at intermediate magnitudes among the
subfields forming a mosaic indicates that the relative photometry
between them is reasonable, reinforcing the evidence mentioned
above. More importantly, on average, there is an excellent agreement
with the counts of \cite{iovino05}, which reach a comparable depth to
the deep \j\k\ strategy.  This can be seen in
Fig.~\ref{fig:counts_avrg} which shows the average counts obtained
taking into account all fields and regions for the $J$- and $K_s$-band
data. In all cases each field contributes only at magnitudes brighter
than their respective 80\% completeness limit, and no completeness
corrections have been attempted. These mean counts show an excellent
agreement with those recently obtained by different authors
\citep{vaisanen00,martini01, iovino05}, being further strong evidence
of the overall quality of the data. The mean counts for the \j- and
\k-bands are given in Tables~\ref{tab:counts_j} and \ref{tab:counts_k}
listing in Col.~1 the center of the magnitude bin in the Vega system,
in Col.~2 the number of galaxies per 0.5~mag and per square degree,
and in Col.~3 the standard deviation among the contributing fields.

\begin{table}
\caption{The derived raw galaxy counts for the \j -band.  The counts
are given per mag per square degree. Each field only contributes to
the bins brighter than their 80\% completeness limit. No completeness
correction has been attempted.}
\label{tab:counts_j}
\begin{center}
\begin{tabular}{rrr}
\hline\hline
Mag & N  & $\sigma_N$ \\
\hline
 10.25 &        0.0 &        0.0\\
 10.75 &        0.0 &        0.0\\
 11.25 &        0.0 &        0.0\\
 11.75 &        0.0 &        0.0\\
 12.25 &        2.9 &       26.9\\
 12.75 &        2.8 &       26.0\\
 13.25 &        2.8 &       26.0\\
 13.75 &       10.9 &       50.8\\
 14.25 &       11.2 &       52.2\\
 14.75 &       47.4 &      110.8\\
 15.25 &       73.7 &      135.3\\
 15.75 &       93.7 &      156.4\\
 16.25 &      283.3 &      255.6\\
 16.75 &      444.5 &      381.0\\
 17.25 &      766.0 &      463.1\\
 17.75 &     1350.6 &      606.4\\
 18.25 &     2059.3 &      746.3\\
 18.75 &     3189.6 &     1083.7\\
 19.25 &     4988.9 &     1731.7\\
 19.75 &     7804.7 &     2398.9\\
 20.25 &    11671.5 &     3354.3\\
 20.75 &    16064.9 &     3381.6\\
 21.25 &    18383.9 &     5991.3\\
 21.75 &    12933.6 &     8897.6\\
 22.25 &    10949.9 &    10678.0\\
\hline
\end{tabular}
\end{center}
\end{table}

\begin{table}
\caption{The derived raw galaxy counts for the \k -band.  The counts
are given per mag per square degree. Each field only contributes to
the bins brighter than their 80\% completeness limit. No completeness
correction has been attempted.}
\label{tab:counts_k}
\begin{center}
\begin{tabular}{rrr}
\hline\hline
Mag & N  & $\sigma_N$ \\
\hline
11.25 &        0.9 &       14.6\\
 11.75 &        4.1 &       33.6\\
 12.25 &        2.1 &       24.2\\
 12.75 &       11.9 &       54.9\\
 13.25 &       20.2 &       77.1\\
 13.75 &       51.8 &      118.7\\
 14.25 &      121.2 &      209.2\\
 14.75 &      224.4 &      254.5\\
 15.25 &      471.6 &      421.6\\
 15.75 &      747.9 &      509.5\\
 16.25 &     1282.1 &      664.6\\
 16.75 &     2144.8 &      932.2\\
 17.25 &     3598.5 &     1366.0\\
 17.75 &     5721.6 &     1832.1\\
 18.25 &     8872.0 &     2904.3\\
 18.75 &    11606.3 &     3772.5\\
 19.25 &    10333.8 &     5998.6\\
 19.75 &     8330.2 &     5731.2\\
 20.25 &     5295.8 &     5289.0\\
\hline
\end{tabular}
\end{center}
\end{table}

\begin{figure*}
\center
\resizebox{0.8\textwidth}{!}{\includegraphics{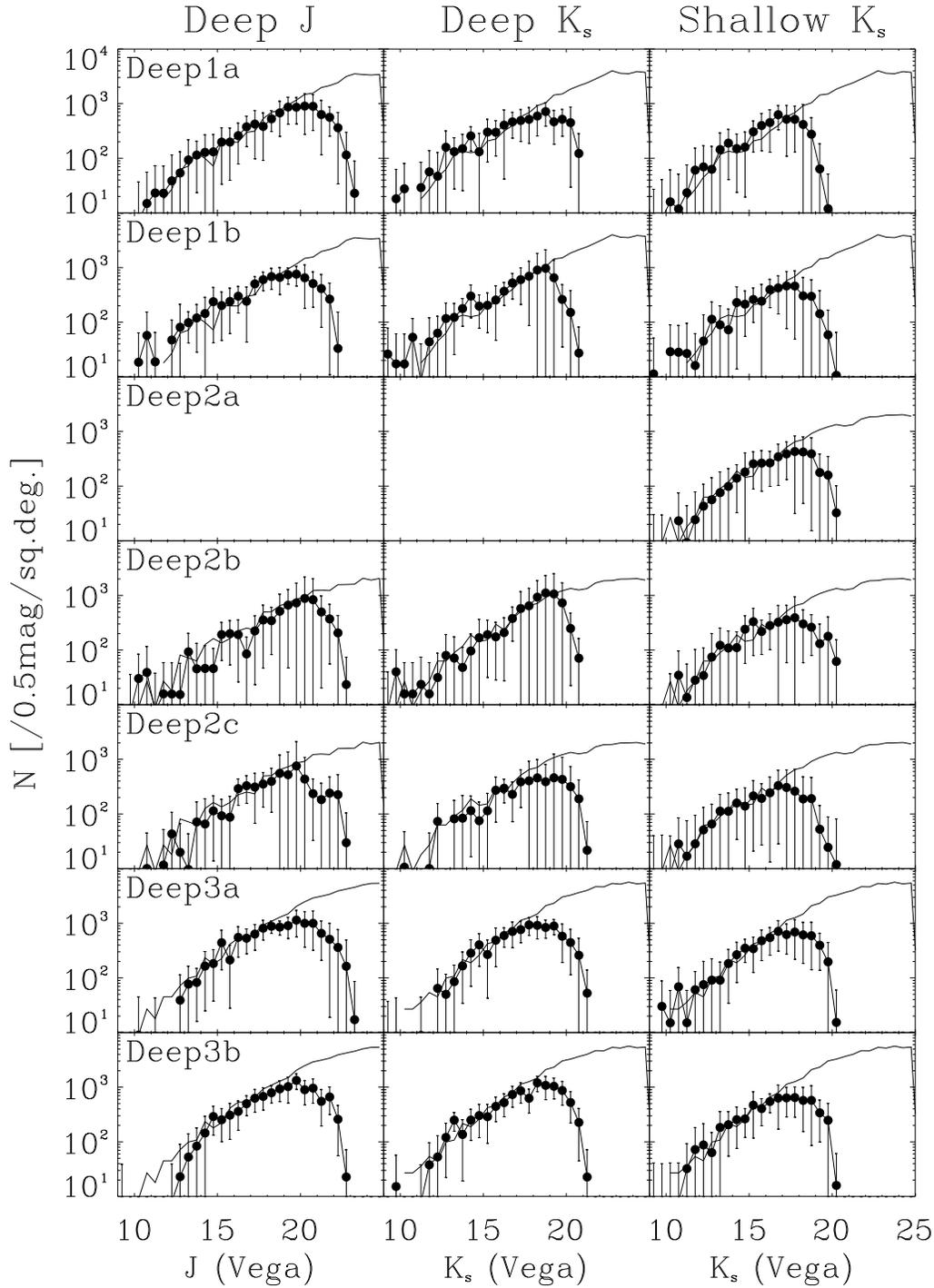}}
\caption{Comparison of star counts (filled circles connected by solid line)  for each mosaic (rows) and strategy 
(columns) compared with model predictions by \citet[ thin line]{girardi05}.  }
\label{fig:starcounts}
\end{figure*}

Finally, the corresponding stellar counts, sources with
CLASS\_STAR$\geq0.9$, are shown in Fig.~\ref{fig:starcounts}. The
counts (solid circles connected by a solid line) are compared to
predictions adopting the model proposed by \citet[thin
lines]{girardi05}. Despite the large scatter due to the small
field-of-view, the agreement is remarkable, considering the different
regions and that the model was parameterized using optical data.

\section{Summary}
\label{sec:summary}

This paper presents the infrared data accumulated by the Deep Public
Survey conducted by the EIS program. The paper presents the
observations, reductions, and some of the main products generated and
administrated by the EIS Data Reduction System which allows the
unsupervised reduction of large amounts of data. For instance, the
data set described in the present paper was reduced overnight using
eight dual-processors computers running Linux.  The highthroughput of
the system also makes it ideal for handling the new generation of
infrared surveys using wide-field cameras. In general, the depth of
the survey is relatively homogeneous and the coverage rather uniform
with exception of the shallow coverage of Deep1.  However,
considerable improvement in coverage and photometric accuracy could
have been reached if the survey was carried out in service rather than
visitor mode allowing a better optimization depending on the actual
weather conditions, in particular ensuring sufficient photometric data
for reliable calibration of all the fields.

The near-infrared dataset being released is one of largest
currently available and its quality has been assessed qualitatively by
visual inspection of the individual images and quantitatively by
direct comparison with measurements made by the 2MASS survey limited
to the bright end and by the comparison of galaxy and star counts
with those of other authors and with model predictions.

The depth and areal coverage in the near-infrared combined with the
deep multi-wavelength optical data reported in Paper~III make this
data set extremely useful to further explore the nature of extremely
red objects \citep[e.g. ][]{kong05} and to search for distant galaxy
clusters. With this paper, all the infrared data accumulated by the EIS
project up to late 2004 are now in the public domain\footnote{The
science grade images and catalogs are available at the CDS via
anonymous ftp to cdsarc.u-strasbg.fr (130.79.128.5) or via
http://cdsweb.u-strasbg.fr/cgi-bin/qcat?J/A+A/}.

\acknowledgements We thank the anonymous referee for useful comments
which improved the manuscript.  This publication makes use of data
products from the Two Micron All Sky Survey, which is a joint project
of the University of Massachusetts and the Infrared Processing and
Analysis Center/California Institute of Technology, funded by the
National Aeronautics and Space Administration and the National Science
Foundation. We thank all of those directly or indirectly involved in
the EIS effort.  Our special thanks to M. Scodeggio for his continuing
assistance, A. Bijaoui for allowing us to use tools developed by him
and collaborators and past EIS team members for building the
foundations of this program.  LFO acknowledges financial support from
the Carlsberg Foundation, the Danish Natural Science Research Council
and the Poincar\'e Fellowship program at Observatoire de la C\^ote
d'Azur. The Dark Cosmology Centre is funded by the Danish National
Research Foundation.  JMM would like to thank the Observatoire de la
C\^{o}te d'Azur for its hospitality during the writing of this paper.

\appendix

\section{Coverage per region}
\label{app:coverage}

In this appendix, figures giving an overview of the final coverage for
each mosaic are presented. Each figure covers one region (Deep1,
Deep2 or Deep3). For Deep1 there are three figures (\j , deep \k , and
shallow
\k ), while for Deep2 and Deep3 only two figures (\j\ and \k ) are
available since here the deep \k\ strategy fills the central parts of
the shallow mosaics. The relative sizes among the fields reflect
their relative sky-coverage.

From the figures it is clear that the areal completeness of the
mosaics are in general high, even though the Deep1 shallow \k\ mosaics
suffer from incompleteness, leading to large difficulties in creating
mosaics and carrying out an internal re-calibration to establish a
common magnitude system for all frames. The centers of each subfield
can be found in Table~\ref{tab:obs}.

\begin{figure*}
\center
\resizebox{0.571\textwidth}{!}{\includegraphics{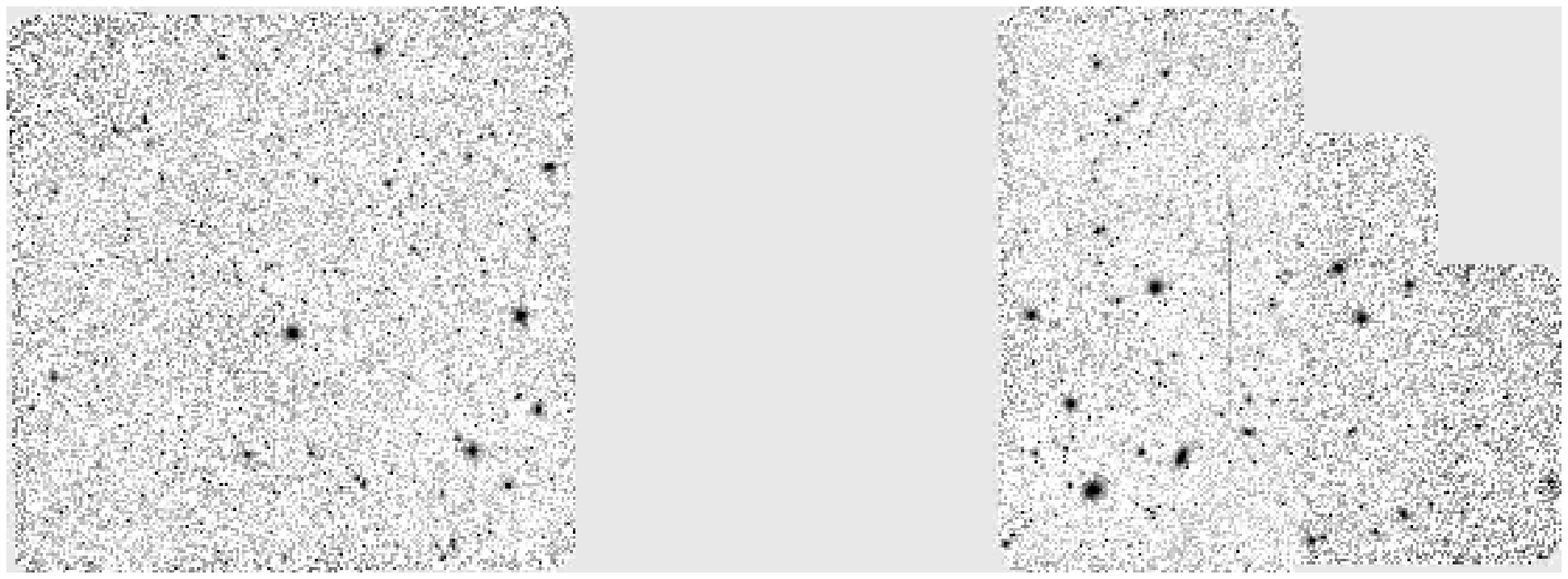}}

\resizebox{0.571\textwidth}{!}{\includegraphics{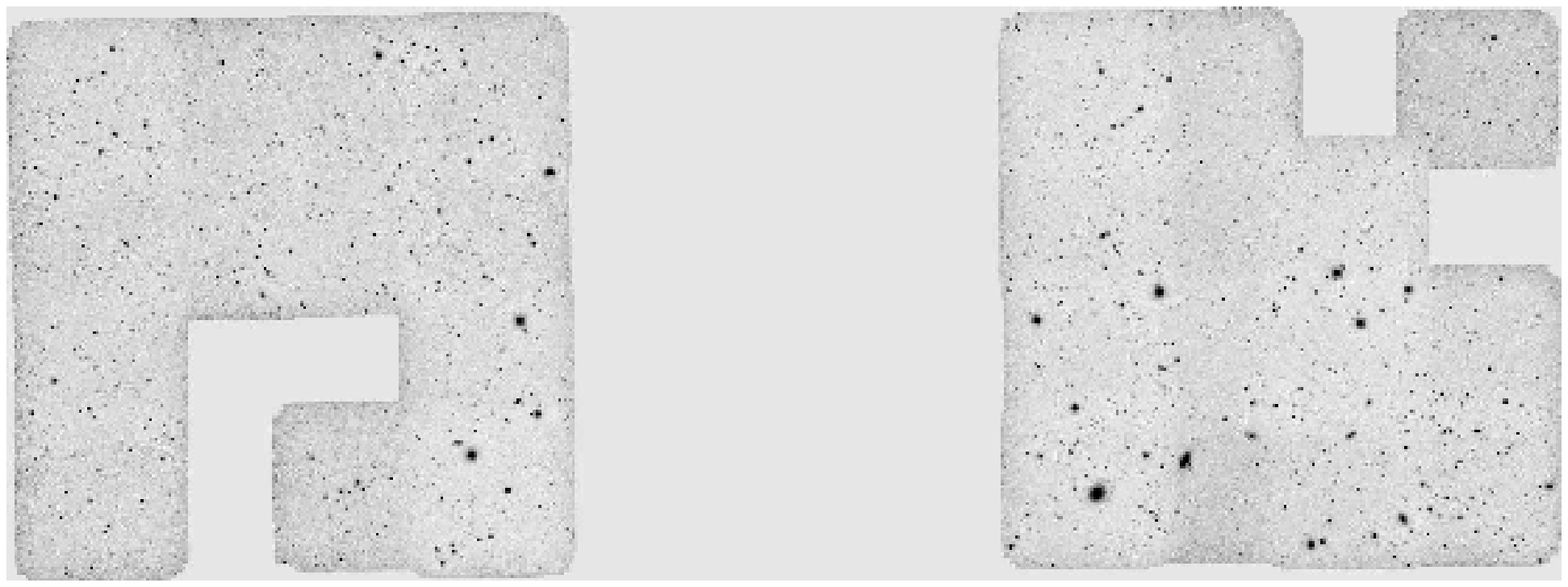}}

\resizebox{0.708\textwidth}{!}{\includegraphics{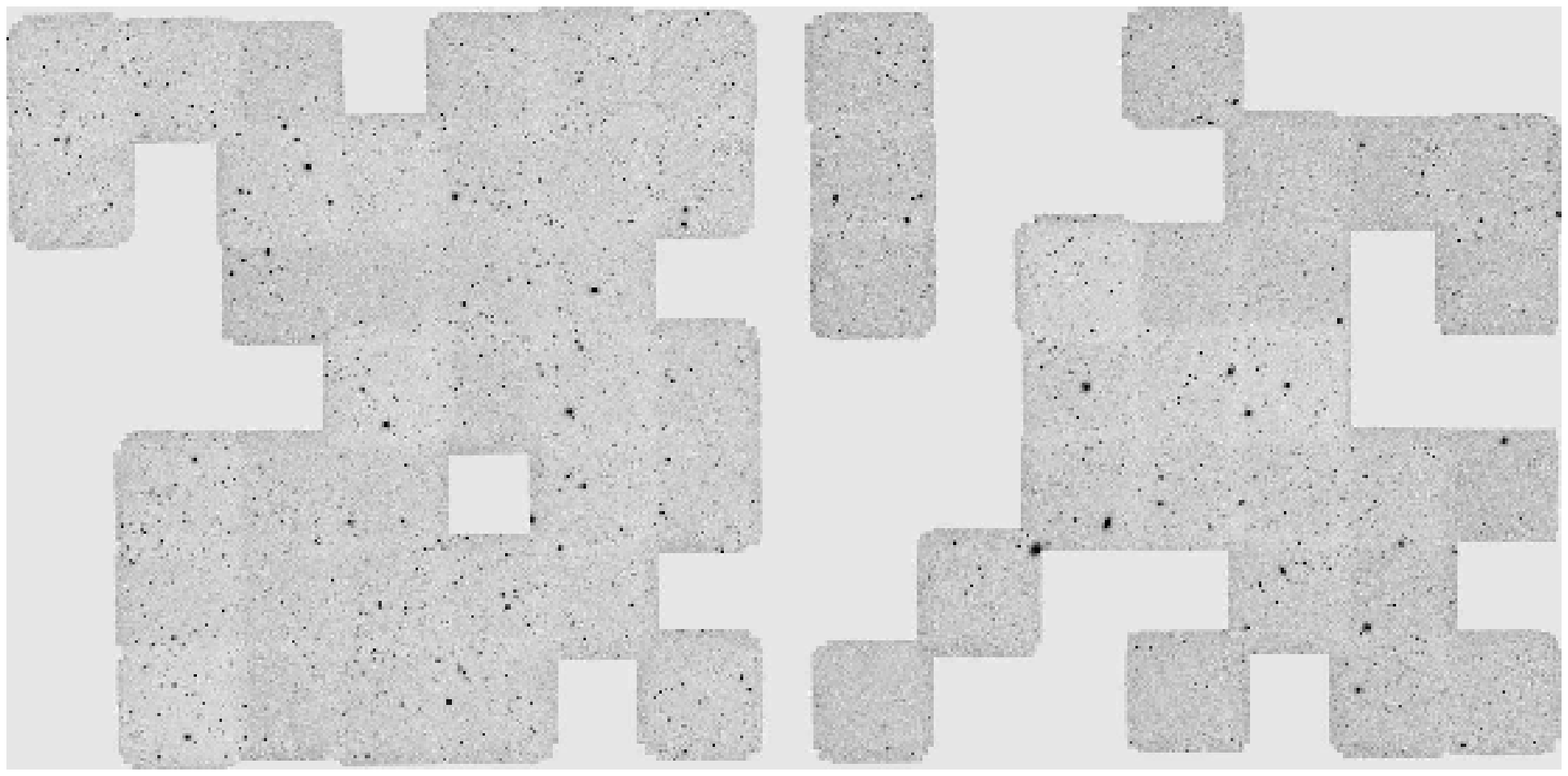}}
\caption{The coverage in the Deep1 region (north up, east left). From
top to bottom it is \j\ ($51\farcm8\times19\farcm0$), deep \k\
($51\farcm9\times19\farcm1$), and shallow \k\
($64\farcm3\times31\farcm5$).  The two fields included are Deep1a
(left) and Deep1b (right).}
\label{fig:deep1_coverage}
\end{figure*}

\begin{figure*}
\center
\resizebox{0.578\textwidth}{!}{\includegraphics{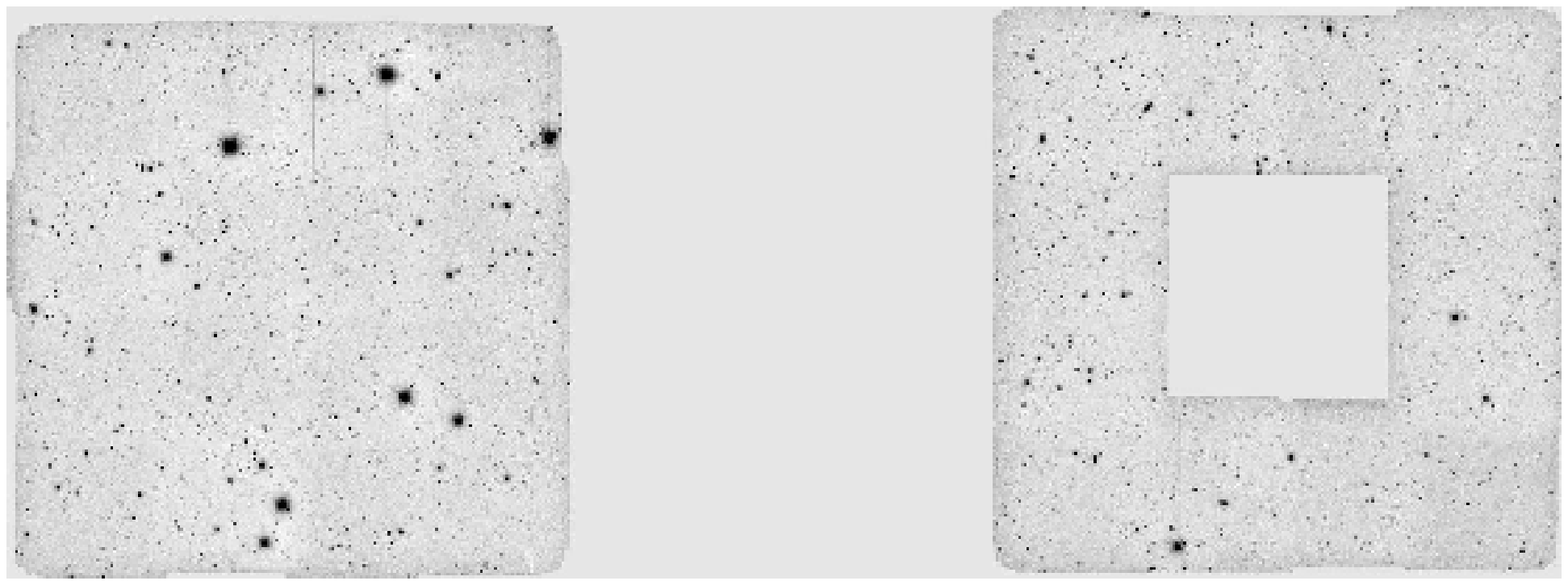}}

\resizebox{\textwidth}{!}{\includegraphics{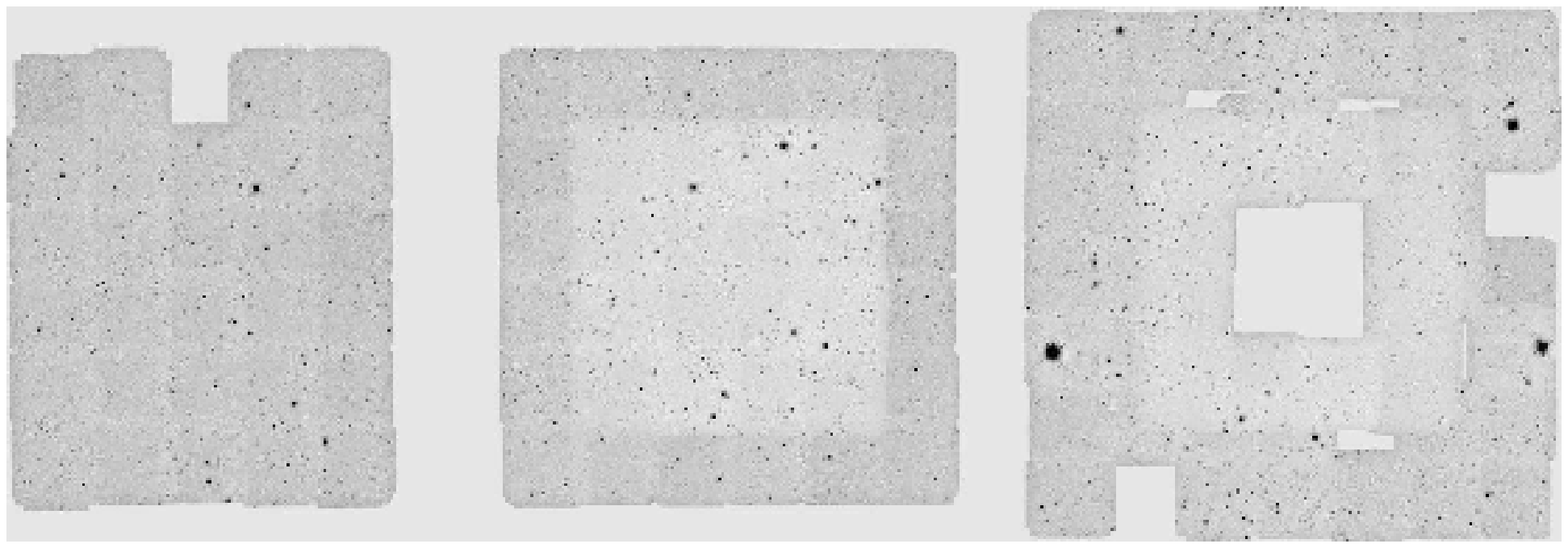}}
\caption{The coverage in the Deep2 region (north up, east left). The
\j\ coverage ($52\farcm5\times19\farcm3$) is on the top and \k\
($90\farcm8\times31\farcm2$) on the bottom.  The fields in the upper
panel are Deep2b (left) and Deep2c (right). The fields in the lower
panel are Deep2a, Deep2b and Deep2c from left to right. In the \k
-band the central parts in \k\ for the mosaics Deep2b and Deep2c are
the deep strategy, for Deep2a only shallow data exists.}
\label{fig:deep2_coverage}
\end{figure*}

\begin{figure*}
\center
\resizebox{0.572\textwidth}{!}{\includegraphics{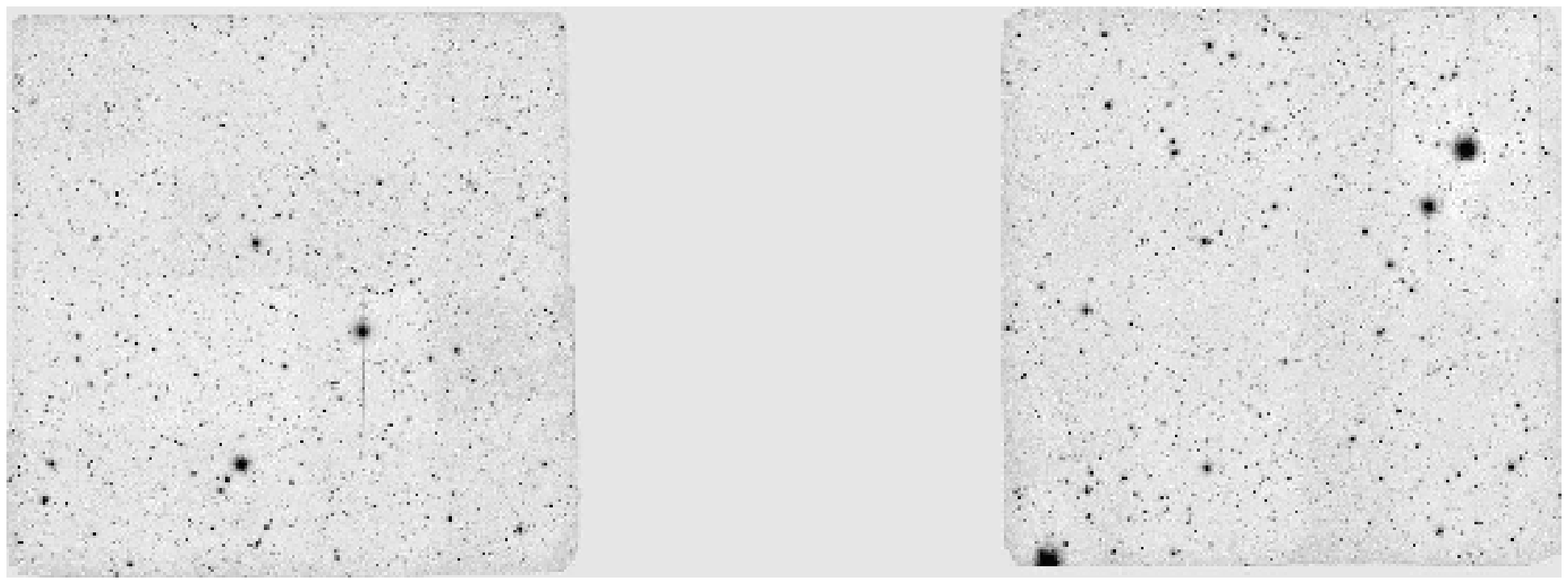}}

\resizebox{0.661\textwidth}{!}{\includegraphics{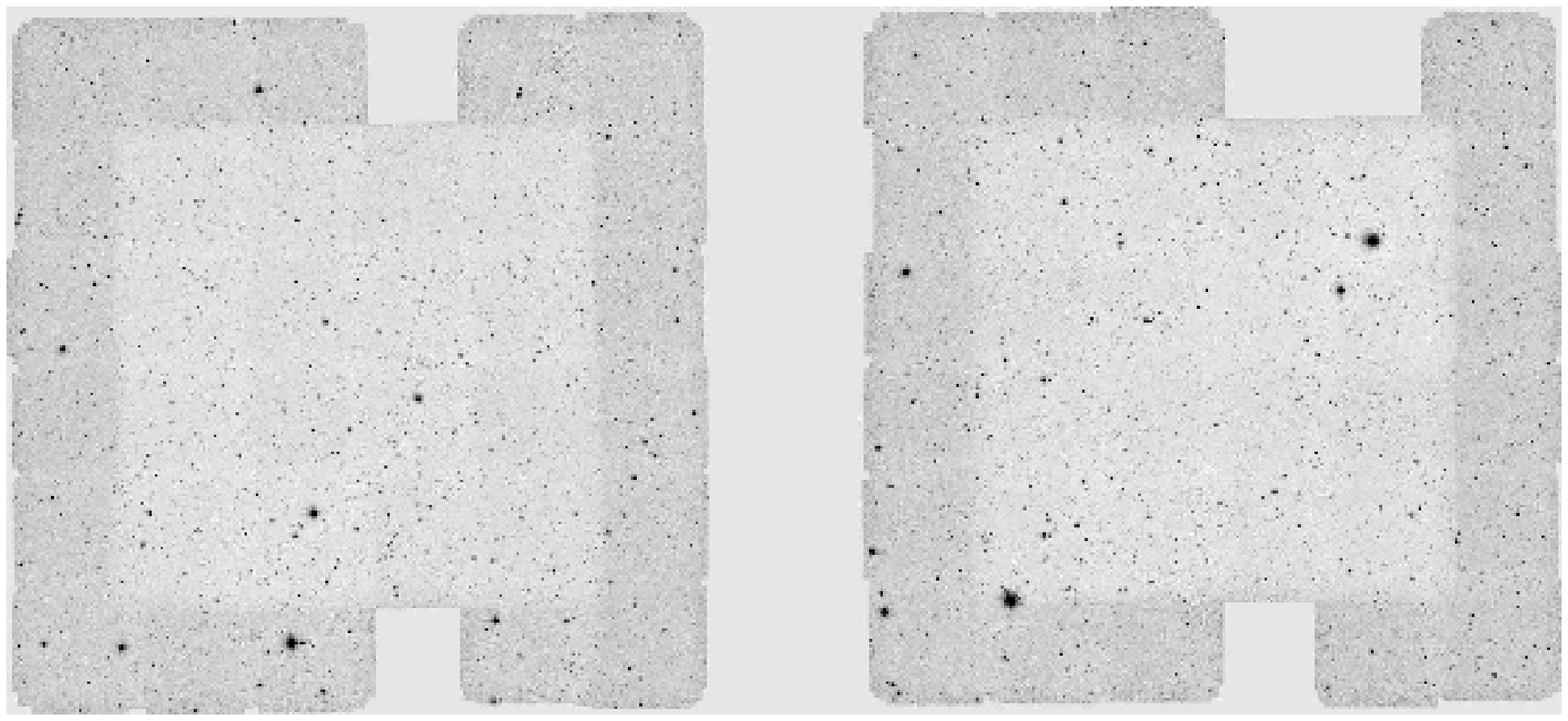}}
\caption{The coverage in the Deep3 region (north up, east left) with
\j\ ($52\farcm0\times19\farcm1$) on the top and \k\
($60\farcm1\times27\farcm3$) on the bottom.  The fields are Deep3a
(left) and Deep3b (right). For the \k -band the central parts
correspond to the deep strategy.}
\label{fig:deep3_coverage}
\end{figure*}

\bibliographystyle{../../../aa}
\bibliography{/home/lisbeth/tex/lisbeth_ref}

\begin{thebibliography}{20}
\expandafter\ifx\csname natexlab\endcsname\relax\def\natexlab#1{#1}\fi

\bibitem[{Cristiani {et~al.}(2000)Cristiani, Appenzeller, Arnouts, Nonino,
  Arag\'on-Salamanca, Benoist, da~Costa, Dennefeld, Rengelink, Renzini,
  Szeifert, \& White}]{Cristiani00}
Cristiani, S., Appenzeller, I., Arnouts, S., {et~al.} 2000, A\&A, 359, 489

\bibitem[{da~Costa {et~al.}(1998)da~Costa, Nonino, Rengelink, Zaggia, Benoist,
  Erben, Wicenec, Scodegiio, Olsen, Guarnieri, Deul, D'Odorico, Hook, Moorwood,
  \& Slijkhuis}]{dacosta98}
da~Costa, L., Nonino, M., Rengelink, R., {et~al.} 1998, astro-ph/9812105

\bibitem[{{da Costa} {et~al.}(2004){da Costa}, {Rite}, \&
  {Slijkhuis}}]{dacosta04}
{da Costa}, L., {Rite}, C., \& {Slijkhuis}, R.~G. 2004, in ASP Conf. Ser. 314:
  Astronomical Data Analysis Software and Systems (ADASS) XIII, 3

\bibitem[{Dickinson \& Giavalisco(2003)}]{dickinson03}
Dickinson, M. \& Giavalisco, M. 2003, in The Mass of Galaxies at Low and High
  Redshift, 324

\bibitem[{Dietrich {et~al.}(2006)Dietrich, Miralles, Olsen, da~Costa, Schwope,
  Benoist, Hambaryan, Mignano, Motch, Rite, Slijkhuis, Tedds, Vandame, Watson,
  \& Zaggia}]{dietrich05}
Dietrich, J.~P., Miralles, J.-M., Olsen, L.~F., {et~al.} 2006, A\&A, 449, 837,
  Paper~I

\bibitem[{Djamdji {et~al.}(1993)Djamdji, Bijaoui, \& Mani\`{e}re}]{djamdji93}
Djamdji, J., Bijaoui, A., \& Mani\`{e}re, R. 1993, in Photogrammetric
  Engineering and Remote Sensing, Vol.~59, 645

\bibitem[{Elston {et~al.}(2006)Elston, Gonzalez, McKenzie, Brodwin, Brown,
  Cardona, Dey, Dickinson, Eisenhardt, Jannuzi, Y.-T, Mohr, Raines, Stanford,
  \& Stern}]{elston05}
Elston, R.~J., Gonzalez, A.~H., McKenzie, E., {et~al.} 2006, ApJ, 639, 816

\bibitem[{{Giavalisco} {et~al.}(2004){Giavalisco}, {Ferguson}, {Koekemoer},
  {Dickinson}, {Alexander}, {Bauer}, {Bergeron}, {Biagetti}, {Brandt},
  {Casertano}, {Cesarsky}, {Chatzichristou}, {Conselice}, {Cristiani}, {Da
  Costa}, {Dahlen}, {de Mello}, {Eisenhardt}, {Erben}, {Fall}, {Fassnacht},
  {Fosbury}, {Fruchter}, {Gardner}, {Grogin}, {Hook}, {Hornschemeier}, {Idzi},
  {Jogee}, {Kretchmer}, {Laidler}, {Lee}, {Livio}, {Lucas}, {Madau},
  {Mobasher}, {Moustakas}, {Nonino}, {Padovani}, {Papovich}, {Park},
  {Ravindranath}, {Renzini}, {Richardson}, {Riess}, {Rosati}, {Schirmer},
  {Schreier}, {Somerville}, {Spinrad}, {Stern}, {Stiavelli}, {Strolger},
  {Urry}, {Vandame}, {Williams}, \& {Wolf}}]{giavalisco04}
{Giavalisco}, M., {Ferguson}, H.~C., {Koekemoer}, A.~M., {et~al.} 2004, ApJ,
  600, L93

\bibitem[{Girardi {et~al.}(2005)Girardi, Groenewegen, Hatziminaoglou, \&
  da~Costa}]{girardi05}
Girardi, L., Groenewegen, M., Hatziminaoglou, E., \& da~Costa, L. 2005, A\&A,
  436, 895

\bibitem[{Iovino {et~al.}(2005)Iovino, McCracken, Garilli, Foucaud, Fèvre,
  Maccagni, Saracco, Bardelli, Busarello, Scodeggio, Zanichelli, Paioro,
  Bottini, Brun, Picat, Scaramella, Tresse, Vettolani, Adami, Arnaboldi,
  Arnouts, Bolzonella, Cappi, Charlot, Ciliegi, Contini, Franzetti, Gavignaud,
  Guzzo, Ilbert, Marano, Marinoni, Mazure, Meneux, Merighi, Paltani, Pell\'{o},
  Pozzetti, Radovich, Zamorani, Zucca, Bertin, Bondi, Bongiorno, Cucciati,
  Gregorini, Mathez, Mellier, Merluzzi, Ripepi, \& Rizzo}]{iovino05}
Iovino, A., McCracken, H., Garilli, B., {et~al.} 2005, A\&A, 442, 423

\bibitem[{Kong {et~al.}(2006)Kong, Daddi, Arimoto, Renzini, Broadhurst,
  Cimatti, Ikuta, Ohta, da~Costa, Olsen, Onodera, \& N.Tamura}]{kong05}
Kong, X., Daddi, E., Arimoto, N., {et~al.} 2006, ApJ, 638, 72

\bibitem[{{Maddox} {et~al.}(1990){Maddox}, {Efstathiou}, \&
  {Sutherland}}]{maddox90}
{Maddox}, S.~J., {Efstathiou}, G., \& {Sutherland}, W.~J. 1990, \mnras, 246,
  433

\bibitem[{Martini(2001)}]{martini01}
Martini, P. 2001, AJ, 121, 598

\bibitem[{Mignano {et~al.}(2006)Mignano, Miralles, da~Costa, Olsen, Prandoni,
  Arnouts, Benoist, Madejsky, Rit\'{e}, Slijkhuis, Vandame, \&
  Zaggia}]{mignano06}
Mignano, A., Miralles, J.-M., da~Costa, L., {et~al.} 2006, submitted to A\&A,
  Paper~III

\bibitem[{Moorwood {et~al.}(1998)Moorwood, Cuby, \& Lidman}]{moorwood98}
Moorwood, A., Cuby, J., \& Lidman, C. 1998, The Messenger, 91

\bibitem[{Olsen {et~al.}(2006)Olsen, Miralles, da~Costa, Benoist, Vandame,
  Rengelink, Rit\'{e}, Scodeggio, Slijkhuis, Wicenec, \& Zaggia}]{olsen06a}
Olsen, L.~F., Miralles, J.-M., da~Costa, L., {et~al.} 2006, A\&A, 452, 119,
  Paper~II

\bibitem[{Persson {et~al.}(1998)Persson, Murphy, Krzeminski, Roth, \&
  Rieke}]{persson98}
Persson, S., Murphy, D., Krzeminski, W., Roth, M., \& Rieke, M. 1998, AJ, 116

\bibitem[{Rengelink {et~al.}(1998)Rengelink, Nonino, da~Costa, Zaggia, Erben,
  Benoist, Wicenec, Scodeggio, Olsen, Guarnieri, Deul, Hook, Moorwood, \&
  Slijkhuis}]{rengelink98}
Rengelink, R., Nonino, M., da~Costa, L., {et~al.} 1998, astro-ph/9812190

\bibitem[{Skrutskie {et~al.}(2006)Skrutskie, Cutri, Stiening, Weinberg,
  Schneider, Carpenter, Beichman, Capps, Chester, Elias, Huchra, Liebert,
  Lonsdale, Monet, Price, Seitzer, Jarrett, Kirkpatrick, Gizis, Howard, Evans,
  Fowler, Fullmer, Hurt, Light, Kopan, Marsh, McCallon, Tam, Dyk, , \&
  Wheelock}]{skrutskie06}
Skrutskie, M.~F., Cutri, R.~M., Stiening, R., {et~al.} 2006, AJ, 131, 1163

\bibitem[{V{\"a}is{\"a}nen {et~al.}(2000)V{\"a}is{\"a}nen, Tollestrup, Willner,
  \& Cohen}]{vaisanen00}
V{\"a}is{\"a}nen, P., Tollestrup, E., Willner, S., \& Cohen, M. 2000, AJ, 540,
  593

\end{thebibliography}

\end{document}